\documentclass[a4paper,11pt]{article}
\pdfoutput=1 

\usepackage{jcappub} 

\usepackage[T1]{fontenc} 
\usepackage{comment}
\usepackage{xspace}
\usepackage{multirow}
\usepackage{indentfirst}

\usepackage[dvipsnames,table]{xcolor}
\definecolor{darkred}{rgb}{0.5,0,0}
\definecolor{darkblue}{rgb}{0,0,0.5}
\definecolor{firebrick}{rgb}{0.75,0.125,0.125}
\definecolor{darkgreen}{rgb}{0,0.5,0}

\newcommand{\dummyfig}[1]{
  \fbox{
    \begin{minipage}[c][0.35\textwidth][c]{0.75\columnwidth}
      \centering{\url{#1}}
    \end{minipage}
  }
}

\newcommand{\dummyfigbi}[1]{
  \centering
  \fbox{
    \begin{minipage}[c][0.4\textwidth][c]{0.45\columnwidth}
      \centering{\url{#1}}
    \end{minipage}
  }
}

\newcommand{\dummyfigtri}[1]{
  \centering
  \fbox{
    \begin{minipage}[c][0.3\textwidth][c]{0.25\columnwidth}
      \centering{\url{#1}}
    \end{minipage}
  }
}

\def\Offline{\mbox{$\overline{\textrm{Off}}$\hspace{.05em}\protect\raisebox{.4ex}{$\protect\underline{\textrm{line}}$}}\xspace}

\newcommand{\Nmu}{N_\mu}

\newcommand{\Xmax}{X_{\rm max}}

\title{\boldmath Sensitivity of Hybrid Particle Detectors to the Energy Spectra of Air Shower Components}

\author[a,b]{P. Assis}
\author[a,b]{R. Concei\c{c}\~ao}
\author[a,b]{P. J. Costa}
\author[a,b]{M. Freitas}
\author[a,b]{B. Tom\'e}

\affiliation[a]{Departamento de F\'isica, Instituto Superior T\'{e}cnico (IST),\\ Universidade de Lisboa, Av. Rovisco Pais 1, 1049-001 Lisbon, Portugal}
\affiliation[b]{Laboratório de Instrumentação e Física Experimental de Partículas (LIP),\\ Av. Prof. Gama Pinto, 2, 1649-003 Lisbon, Portugal}

\abstract{Measuring the energy spectrum of air shower components crucial for understanding primary cosmic rays and the physical processes governing their interactions in the atmosphere. However, accurately measuring the energy of shower particles reaching the ground is challenging due to the inherent simplicity of typical cosmic ray experiments.

This study takes advantage of a hybrid detector station and introduces an analysis for assessing the energy spectra of air shower components. The station is composed of a scintillator surface detector (SSD), a water-Cherenkov detector (WCD), and resistive plate chambers (RPCs). Our results demonstrate that the combination of data from these detectors enables precise assessment of the energy spectrum of electromagnetic particles and low-energy muons, independent of potential detector ageing effects.}

\begin{document}
\maketitle
\flushbottom

\section{Introduction}
\label{sec:Intro}

Ultra-high-energy cosmic rays (UHECRs) are particles of extraterrestrial origin with centre-of-mass energies exceeding those achieved in state-of-the-art accelerators. The study of UHECRs offers a unique opportunity to probe hadronic interaction models in previously unexplored regions of phase space, while also providing clues about their astrophysical sources and acceleration mechanisms.

When UHECRs interact with the Earth's atmosphere, they initiate extensive air showers (EAS), which serve as indirect probes of the primary particle’s properties. Key observables, such as the number of muons at the ground level ($\Nmu$) and the atmospheric depth of maximum shower development ($\Xmax$), are sensitive to the energy and mass composition of the primary particle that initiated the shower. However, their interpretation is inherently dependent on hadronic interaction models, which are constrained by accelerator data but require extrapolation to ultra-high energies, leading to uncertainties.

Recent results from the Pierre Auger Observatory have demonstrated that state-of-the-art hadronic interaction models fail to describe the observed extensive air showers consistently, regardless of the assumed primary mass composition~\cite{PierreAuger:2024neu}. In particular, simulations exhibit a persistent deficit in the predicted number of muons compared to experimental data~\cite{PierreAuger:2021qsd,whispMuonDeficit}, a discrepancy commonly referred to as the \emph{EAS Muon Puzzle}.

Addressing this discrepancy requires a deeper understanding of the underlying physical mechanisms governing air shower development. This effort is being pursued through experimental upgrades, such as \emph{AugerPrime}~\cite{Castellina:2019irv}, aimed at enhancing the characterization of shower components. Additionally, dedicated accelerator experiments, such as the planned proton-oxygen collisions at the Large Hadron Collider~\cite{pO_coll_LHC}, will provide crucial constraints on hadronic interaction models at energies relevant to ultra-high-energy cosmic-ray secondary interactions.

The extensive number of interactions and particles produced during the development of a shower gives rise to a phenomenon known as \emph{shower universality}. This principle states that key shower characteristics, such as the secondary particle energy spectrum and the evolution of the longitudinal profile, can be entirely determined given the shower development stage and primary energy — independent of the primary mass composition or the specific details of hadronic interactions. This universality has been observed in EAS simulations for both the electromagnetic~\cite{EM_all_2,Śmiałkowski_2018,GILLER201592, Nerling_2006, Giller:2004cf, Conceicao:2015toa, LIPARI2009309} and muonic~\cite{muonUniv1, muonUniv2,muonUniv3, Cazon_2023} components.

A more detailed investigation of shower universality in the muonic sector suggests that hadronic interaction models can be solely differentiated by analyzing the muon energy spectrum~\cite{muonUniv}. Moreover, while the average shape of the electromagnetic longitudinal profile has been found to be consistent with simulation expectations~\cite{PierreAuger:2018gfc}, to our knowledge, its energy spectrum has never been systematically measured. The absence of such measurements is primarily due to the inherent design constraints of cosmic ray detectors, which prioritize simplicity and resilience in harsh observational conditions to cover large areas.

In this work, we propose a methodology that utilizes the analysis of a single \emph{multi-hybrid} detector station to evaluate the energy spectrum of electromagnetic and muonic secondary particles in extensive air showers. The paper is organized as follows: The first section provides an introduction and motivation for the study. Section~\ref{sec:Method} outlines the core concepts and the strategy employed to extract the energy spectrum of secondary shower particles, along with a description of the simulation framework. Section~\ref{sec:Results} presents and discusses the obtained results for the electromagnetic and muonic energy spectra. Section~\ref{sec:Exp} explores key experimental considerations for planning a measurement based on this analysis, including energy and angular resolution in shower reconstruction, the accuracy of shower core determination, and the impact of detector ageing. The paper concludes with a summary of the main findings.

\section{Principle for assessing the energy spectrum of shower particles}
\label{sec:Method}

The primary strategy of this paper is to leverage the distinct response characteristics of a hybrid detector station by combining the outputs of different detectors to infer the energy spectrum of particles that simultaneously traverse all detection systems. To achieve this goal, it is advantageous to have one detector that is relatively insensitive to particle energy and another one that exhibits high sensitivity to it. For the former, a scintillation-based detector serves as a suitable choice, while for the latter, we select a water-Cherenkov detector (WCD). 
Incorporating a third detector is desirable to mitigate potential calibration issues. In this work, we choose to position a resistive plate chamber (RPC) beneath the WCD to provide an additional independent measurement.

\begin{figure}[ht]
 \centering
 \includegraphics[width=.5\textwidth]{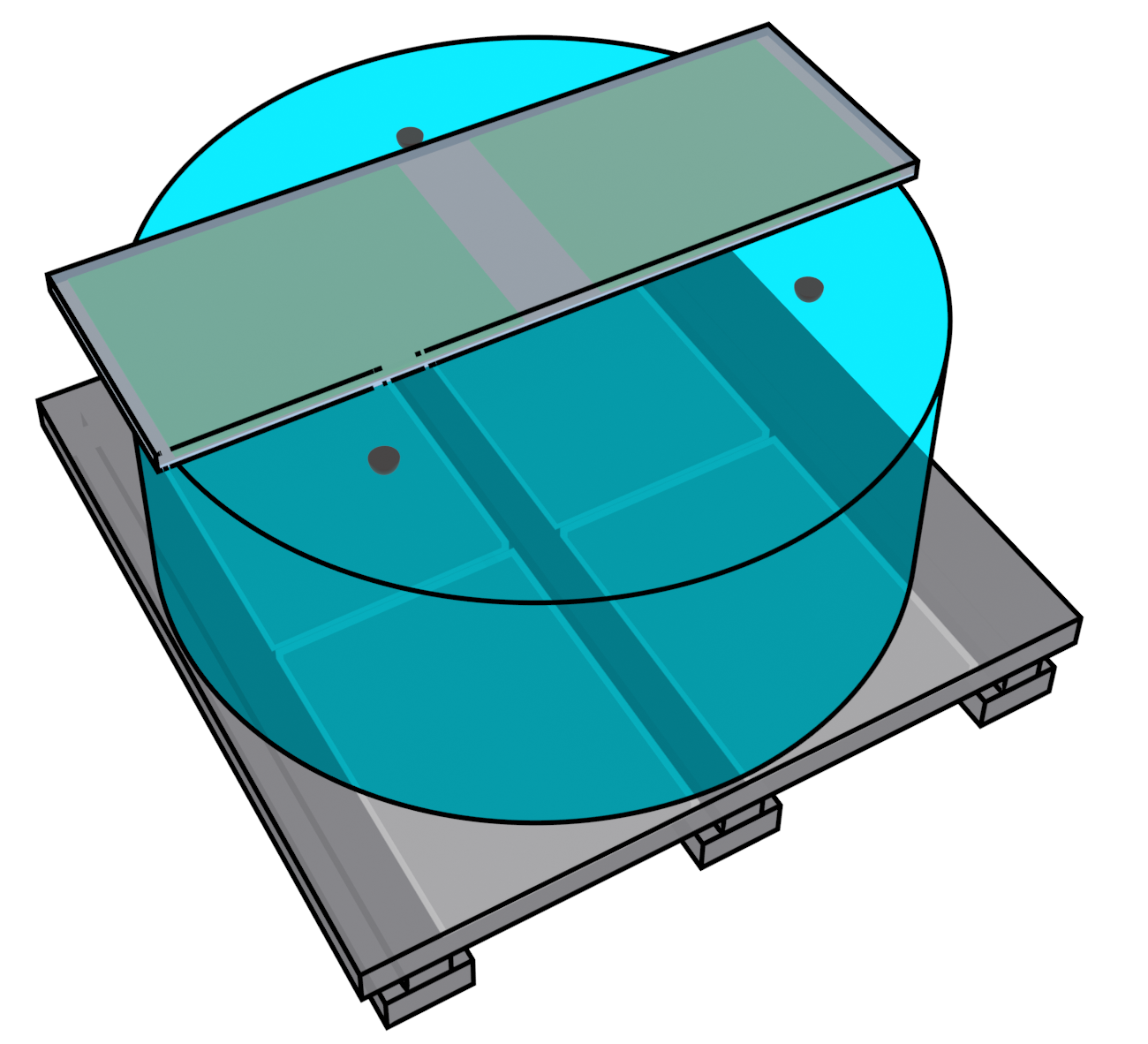}

\caption{A schematic representation of the detector station used in this work.}
\label{fig:detector_scheme}
\end{figure}

Hence, the detector system in this work consists of a surface scintillation detector (SSD) positioned atop a water-Cherenkov detector, with a resistive plate chamber placed below the WCD, thus allowing shower secondary particles to traverse all three detectors. The dimensions of the considered detectors are as follows: an SSD measuring $1.6 \times 1.2\,{\rm m^2}$, a cylindrical WCD with a surface area of $10\,{\rm m^2}$ and a height of $1.2$ m, and four RPCs, each measuring $1.5 \times 1.2\,{\rm m^2}$. A schematic representation of the station is shown in Fig.~\ref{fig:detector_scheme}. While the specific detector technologies are not crucial as long as the above criteria are met, we evaluate this configuration because it is currently being tested at the Pierre Auger Observatory, utilizing the \emph{AugerPrime} scintillator installed on the top of the MARTA R\&D prototype station~\cite{MARTA}.

As previously stated, particles traversing the scintillator are predominantly minimum ionizing particles (MIPs). Consequently, these detectors primarily function as counters, being mainly sensitive to the number of particles crossing them. Due to their relative insensitivity to the highly abundant secondary photons in the shower, these detectors effectively sample the number of electrons and positrons, which are significantly more prevalent than muons and other particle species.

Water-Cherenkov detectors, on the other hand, detect Cherenkov photons emitted by charged particles traversing the water at velocities exceeding the speed of light in the medium. Secondary shower photons undergo conversion within the WCD, initiating electromagnetic showers that, in turn, generate additional Cherenkov light. Notably, the intensity of Cherenkov light emission is proportional to the size of the electromagnetic shower and, consequently, to the energy of the incoming particle. For non-relativistic muons ($E \lesssim 1\,{\rm GeV}$), the detected signal depends on the particle's velocity, which in this energy range is effectively a measure of its energy.

Since the resistive plate chambers (RPCs) are positioned beneath the WCD, they can be struck either directly by muons that traverse the entire detectors and reach the RPC pads or by secondary particles from electromagnetic showers initiated within the WCD. Additionally, they can register signals from delta rays generated by muons as they pass through the detectors.

\begin{figure}[ht]
 \centering
 \includegraphics[width=1\textwidth]{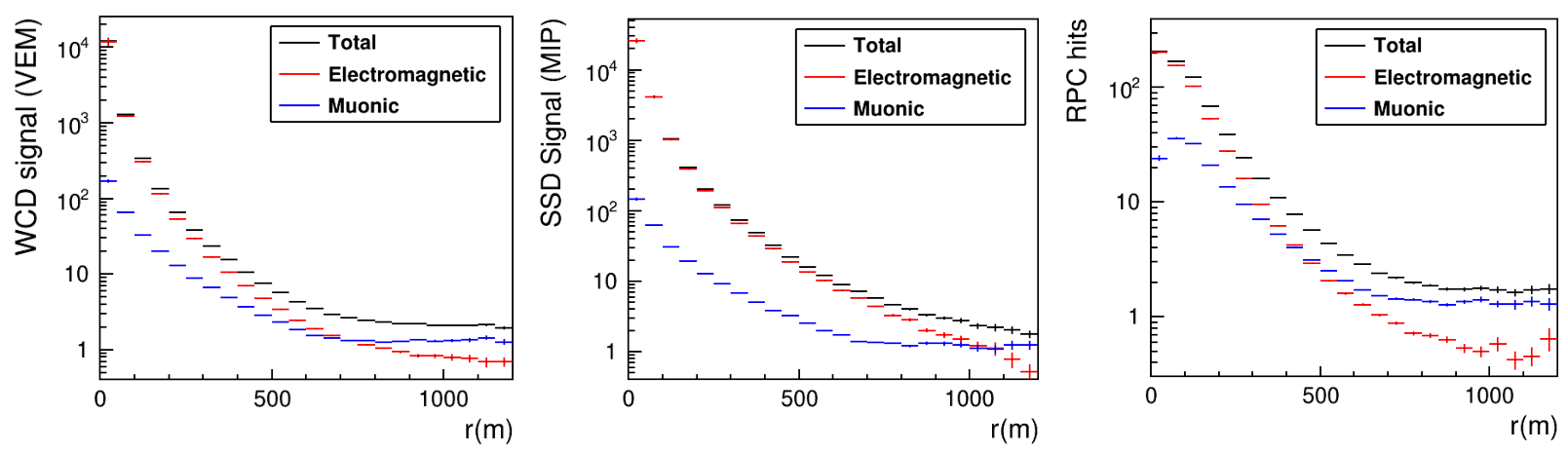}
 \caption{\label{fig:ldfcomp} Average lateral distribution functions (LDFs) for the total (black), electromagnetic (red) and muonic (blue) WCD signals in vertical equivalent muons (VEM) (left), SSD signals in MIP (middle) and RPC particle hits (right). }
\end{figure}

The analysis should be conducted at the station level rather than at the shower event level to fully leverage the multi-hybrid station capabilities. This approach allows for a separate assessment of the electromagnetic and muonic shower components by selecting stations based on their proximity to the shower core.  As shown in Fig.~\ref{fig:ldfcomp}, the contribution of electromagnetic particles and muons to each detector's signal strongly depends on the station's proximity to the shower core. The signal of stations closer to the core is predominantly determined by the electromagnetic component of the shower, whereas at larger distances, the signal is primarily dominated by muons.

As such, in this work, we select two fixed distances to evaluate the contributions of the electromagnetic and muonic shower components to the signal, namely $320\,{\rm m}$ and $700\,{\rm m}$ from the shower core. This multi-hybrid station is embedded within a surface detector array, where stations are spaced $750\,{\rm m}$ apart. This choice is two-fold: first, it corresponds to the array in which this prototype station is currently deployed at the Pierre Auger Observatory; second, this array predominantly detects cosmic rays with energies of $\mathcal{O}(10^{17}\,{\rm eV})$, equivalent to the centre-of-mass energy of proton-proton collisions at $\sqrt{s}=14\,{\rm TeV}$, presently being explored at the Large Hadron Collider~\cite{Slupecki:2888741}, making this measurement even more appealing.

It is important to emphasize that this article is focused on presenting the method for evaluating the energy spectra of shower components at the ground level, and no actual data analysis is included. Any such analysis will likely be addressed in a future publication by the Pierre Auger Collaboration.

\subsection{Simulation Framework}
In this work, the extensive air showers were simulated using CORSIKA v7.7410~\cite{CORSIKA}, and the detector response is obtained using a stand-alone program based on the Geant4 toolkit (v10.5.1)~\cite{Geant4_2006, Geant4_2016}. This standalone simulation was qualitatively validated against the Auger \Offline simulation framework~\cite{Offline} for the three detector responses (SSD, WCD, and RPC), with the detector signals found to be in reasonable agreement. We used a reference configuration for the initial set of simulations, along with additional datasets to assess the robustness of the method employed here with respect to the reconstruction of key shower parameters, such as energy, zenith angle, and core position.

The reference configuration consists of CORSIKA proton-induced air showers with a fixed primary energy of  E $= 10^{17.5} $ eV and a zenith angle of $ \theta = 30^\circ $. A thinning algorithm with $\varepsilon = 10^{-6}$ was applied and hadronic interactions were handled by FLUKA and EPOS-LHC at low and high energies, respectively. In each event, the shower core is injected at a fixed position relative to a specific detector station. This setup serves as a control configuration for our analysis.

To evaluate the dependence of the reconstruction on specific variables, we generated dedicated simulation sets where one parameter was varied at a time. For energy resolution studies, 5000 showers were simulated with energies ranging from  $E = 10^{17.4}$ eV to $E = 10^{17.6}\,$eV. For angular resolution, 2000 showers were generated at a zenith angle of  $ \theta = 50^\circ $. The shower core position was randomly distributed within an area of  $100 \times 100\,{\rm m^2}$ to investigate the impact of core position reconstruction.

The Geant4-based framework used in this study allows users to inject individual particles or CORSIKA-generated air showers into a predefined station configuration. It is also capable of handling thinned CORSIKA EASs, enabling the simulation of higher-energy cascades efficiently.

The core position of these showers can either be set manually or randomized within a specified range. All stations share a common configuration, allowing the placement of a scintillator atop the station and RPCs above and/or below it. Additional station-level customization options include defining the dimensions of the WCD, configuring the layout and size of the photomultiplier tubes (PMTs), and selecting between a single- or double-layer station.

For this study, the WCD dimensions match those of an Auger station, as previously described. A scintillator is placed on top of the station, and an RPC setup is positioned beneath the tank. Inside the tank, three downward-facing hemispherical PMTs are arranged in a triangular pattern, each with a diameter of 13 cm. A Minimum Ionizing Particle (MIP) corresponds to 25 photoelectrons (p.e.), while a Vertical Equivalent Muon (VEM) is set at 84 p.e.

The surface scintillator detector consists of two sets of 24 extruded polystyrene-based scintillating bars, each measuring 1.6 m in length, $5\,$cm in width, and $1\,$cm in thickness. These bars are enclosed within a light-tight aluminium box measuring $3.8\,$m in width and $1.3\,$m in length. The number of photoelectrons registered per hit is parameterized as a function of the hit coordinates along a bar and the energy deposited, accounting for the light attenuation length in the fibres. 

The simulation also includes RPCs, each with a width of approximately $1.1\,$cm in thickness, divided into 64 square pads of equal area, following what is described in~\cite{MARTA}. Each RPC is enclosed within an aluminium box measuring $1.25\,$m in length, $1.65\,$m in width, and $5\,$cm in thickness. These units are placed in designated slots, 15 cm above the ground, within a concrete support structure that supports the WCD. The top slab of this concrete structure, which separates the RPCs from the WCD, is a $15\,$cm thick square with a side length of $3.64\,$m.

A simple threshold-based station-level trigger is implemented, where a station is considered active if all PMTs individually register a signal of at least $1.75\,$
VEM~\cite{PierreAuger:2015eyc}. 

The assessment of the sensitivity of the shower particles' energy spectrum is done in Section~\ref{sec:Results} using the simulated shower energy and geometry. The dependence on the shower reconstructed quantities, such as the core position, direction and energy, is evaluated in Section~\ref{sec:Exp} using dedicated simulation datasets.


\subsection{Method to evaluate the sensitivity to the particle energy spectrum}

In this work, we take the energy spectra of the electromagnetic and muonic components at ground level—obtained from CORSIKA simulations—and modify them while preserving the functional forms of their shapes. This approach is based on the assumption that differences in the energy spectra may arise from the limited ability of hadronic interaction models to accurately describe the underlying physics of shower development, rather than from the fundamental mechanisms of the shower itself. By doing so, we are able to investigate how variations in the particle energy distribution influence the detector response. In particular, we adjust the slopes of the low- and high-energy power-law tails while keeping the total particle multiplicity fixed.

\begin{figure}[ht]
 \centering
 \includegraphics[width=0.49\textwidth]{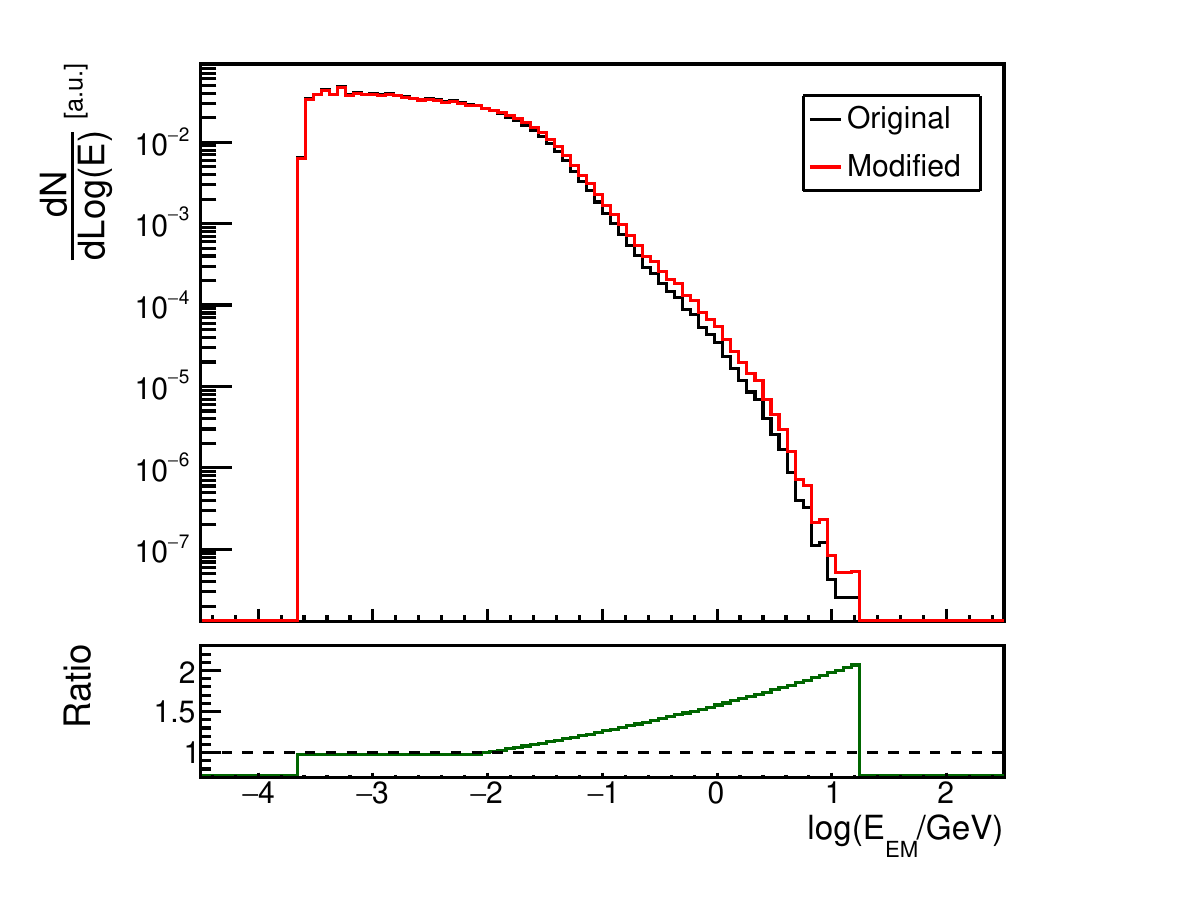}
 \includegraphics[width=0.49\textwidth]{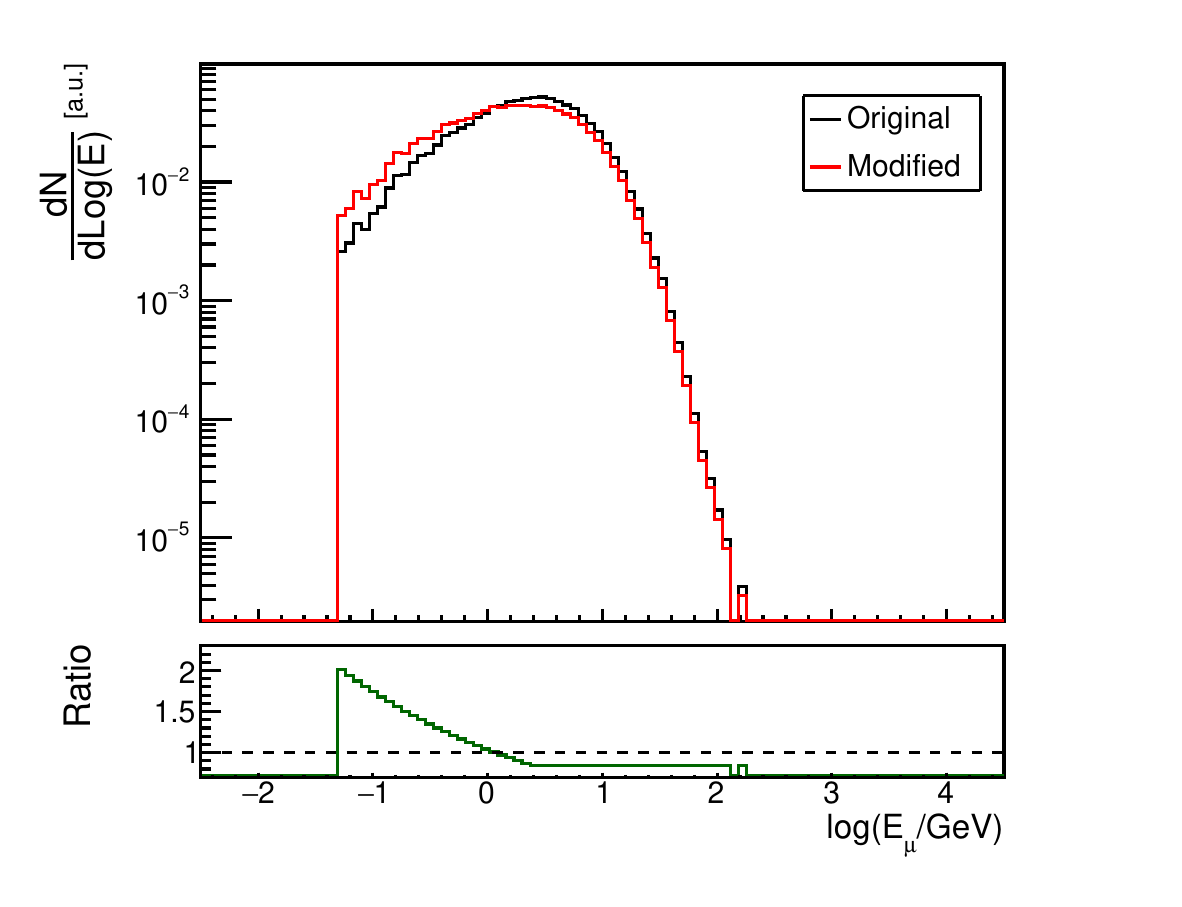}
 \caption{\label{fig:spectra} Examples of modified energy distributions of shower secondary particles for the electromagnetic component (left) and muonic component (right).}
\end{figure}

The original particle energy distribution is obtained by processing CORSIKA-generated air showers down to the detector level using the aforementioned Geant4-based framework. Figure~\ref{fig:spectra} illustrates examples of these distributions for the electromagnetic and muonic shower components. 

To modify the shape of the distributions, we apply an energy-dependent weight in a selected energy range. The weighting function is defined as

\begin{equation}
g(E) = \exp{\bigg(\frac{\ln(c)}{b-a}\cdot (E-a)\bigg)}
\label{eq:modfunc}
\end{equation}

where $E$ is the particle energy, and $a$ and $b$ define the lower and upper bounds of the energy window over which the modification is applied. The parameter $c$ controls the total weight increase (or decrease) across this window. This function modifies the spectrum's tail in a controlled manner. Specifically, it ensures that $g(E) = 1$ at $E = a$, and $g(E) = c$ at $E = b$, resulting in a smooth transition of the weights from 1 to $c$ within a selected energy range. 

The choice of a monotonically increasing function for weighting the original spectrum is deliberate. Applying a flat increase within a limited region would introduce abrupt spikes at the transition between the modified and unmodified regions. This functional form ensures a smooth transition between the modified and unmodified regions, preserving the spectrum's overall shape and features. After applying the weights, the full distribution is renormalized to conserve the total number of particles. 

The second step involves updating the original energy of the particles by sampling from the modified spectrum. By treating the modified spectrum as a probability distribution, random values can be sampled and assigned as new particle energies. Applying this process to all particles ensures that the entire particle ensemble conforms to the modified distribution, thus influencing the detector signals accordingly.

\begin{figure}[ht]
 \centering
 \includegraphics[width=0.49\textwidth]{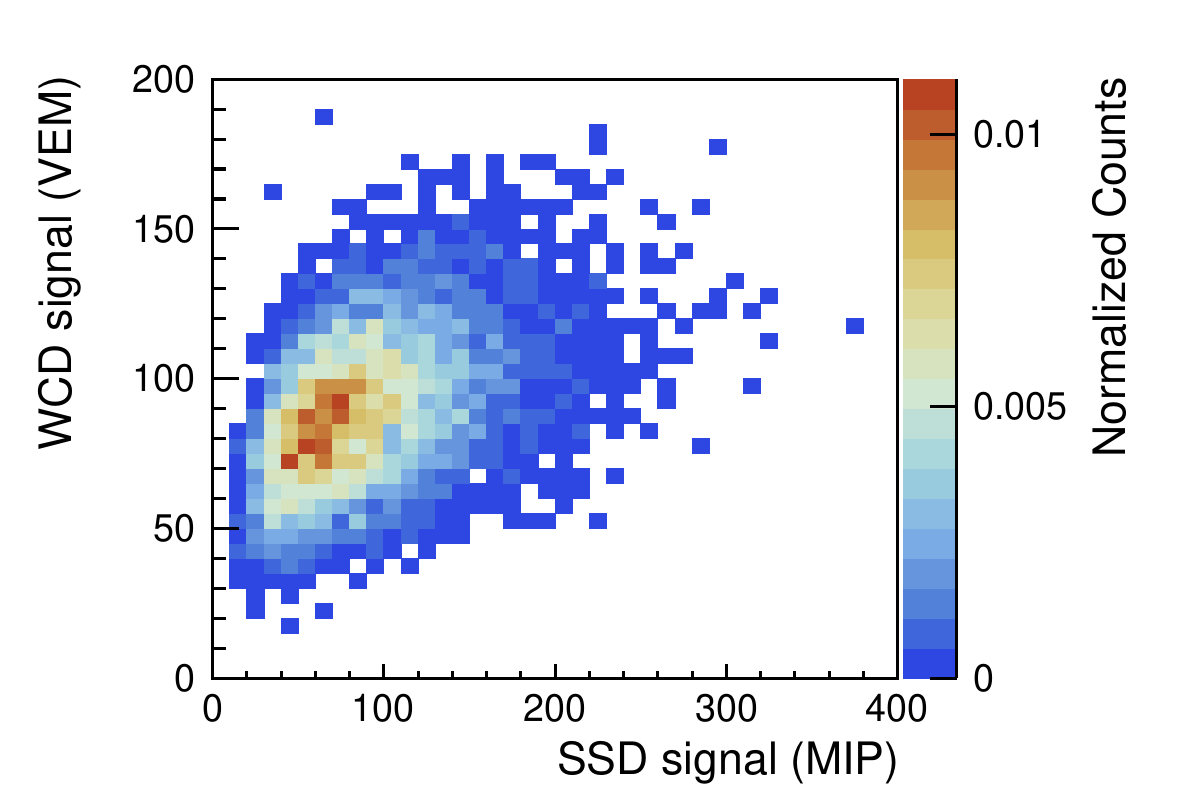}
 \includegraphics[width=0.49\textwidth]{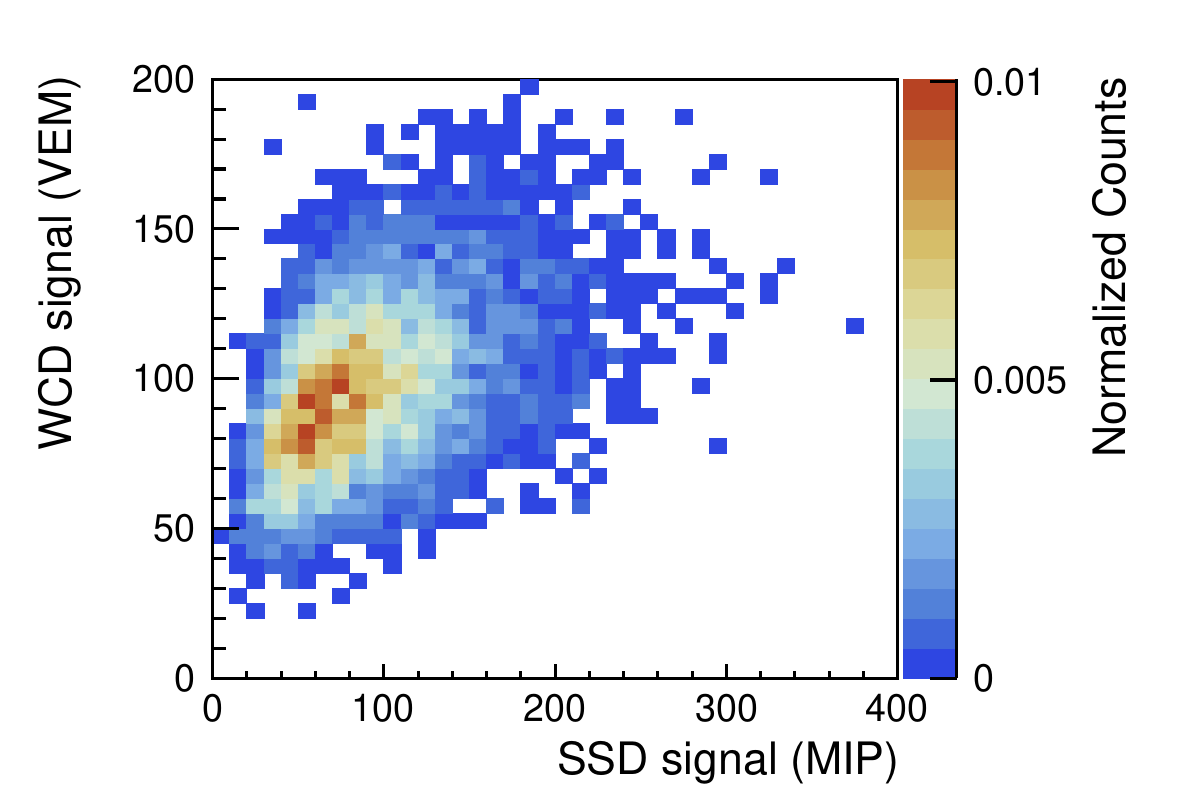}
 \caption{\label{fig:calibplots}Calibration plot between the SSD and WCD signals corresponding to the original (left) and modified (right) particle energy distributions in the station fixed at $r_{\rm core} = 320\,$m.}
\end{figure}

In Fig.~\ref{fig:calibplots}, the correlation between the signals recorded by the scintillator and the WCD is shown for a station located at a distance of $320\,$m from the shower core, $r_{\rm core}$. The left panel presents the detector response population  before modifying the electromagnetic energy spectrum, referred to as \emph{original} in the caption of Fig.~\ref{fig:calibplots} (left). The right panel of Fig.~\ref{fig:calibplots}, displays the correlation after applying the modifications to the energy spectrum, following the histogram labelled \emph{modified} in Fig.~\ref{fig:spectra} (left).

To emphasize the changes in the correlation plot, i.e., the \emph{movement} of the signal response populations following the modification of the energy spectrum, the difference between the right and left histograms in Fig.~\ref{fig:calibplots} is displayed, resulting, for instance, in Fig.~\ref{fig:emSSDvsWCD} (left). It is important to note that the histograms in Fig.~\ref{fig:calibplots} were previously normalized such that their sum equals one.

To enable quantitative comparison across different detector types and avoid unit inconsistencies, the detector signals in the difference plots are normalized to their respective mean values from the original distributions. This yields dimensionless variables $ R_{SSD}, R_{WCD}$ and $R_{RPC}$, which represent the relative change in detector response and allow for direct comparison of sensitivity across different detector technologies.
 
The qualitative differences in the relative difference plots are clearly visible and discussed in the next section (e.g., Fig.\ref{fig:emSSDvsWCD} (left)). Additionally, to perform a quantitative assessment, we introduced two estimators, $R$ and $\theta$, which are derived from these plots and conveniently represented in polar graphs, such as in Fig.\ref{fig:PolarEMhigh}. Both estimators are computed using the barycentres of the positive and negative bin sets in the relative difference plot. The estimator $R$ is defined as the arithmetic distance between the barycentres, while $\theta$ represents the angle, with respect to the $x$-axis, of the vector formed between the barycentres, where the origin of the vector is set at the negative barycentre. The schemes in Fig. \ref{fig:emSSDvsWCD} (left) and Fig. \ref{fig:emWCDvsRPC} (left) provide a geometric interpretation of $R$ and $\theta$, for the SSD-WCD and WCD-RPC difference plots, respectively.

While $R$ and $\theta$ are not direct measurements of the spectrum itself, they provide a useful way to compare experimental data with simulations incorporating specific spectral modifications . By computing these variables from data and comparing them to those from simulated scenarios, we gain insight into the spectral changes that best reproduce the observed detector signals. In this way, $R$ and $\theta$ serve as effective intermediaries between the detector observables and the underlying energy spectra of the shower components, offering a practical method to interpret the data in terms of deviations from the reference simulation. 

\section{Results}
\label{sec:Results}

In this section, the tails of the energy spectrum—both low- and high-energy, as well as electromagnetic and muonic—will be modified for stations at varying distances from the shower core. The resulting changes in the relative difference correlation plot and the estimators introduced in the previous section will be analyzed and discussed.

\subsection{High-energy tail of the electromagnetic component}
\label{subsec:emHEresults}

The energy distribution of electromagnetic particles features a low-energy plateau followed by a high-energy power-law tail. This section presents the results of modifying the high-energy tail of this distribution to slightly enhance its relative contribution and analyze its impact on detector responses.  

For this configuration, the station is positioned at a fixed distance of $r_{\rm core} = 320\,$m to enhance the contribution of electromagnetic particles with respect to muons. The parameters $a = 8\,$MeV and $b = 16\,$GeV were chosen to encompass the entire high-energy tail. As previously noted, the magnitude of the modification scales with the energy of the distribution. The parameter $c$ regulates the overall intensity of the modification and is defined as the maximum ratio between the modified and original distributions. For instance, in the left panel of Figure \ref{fig:spectra}, $c = 2$, corresponding to the ratio observed in the highest energy bin of the spectrum.

\begin{figure}[ht]
 \centering
 \includegraphics[width=0.49\linewidth]{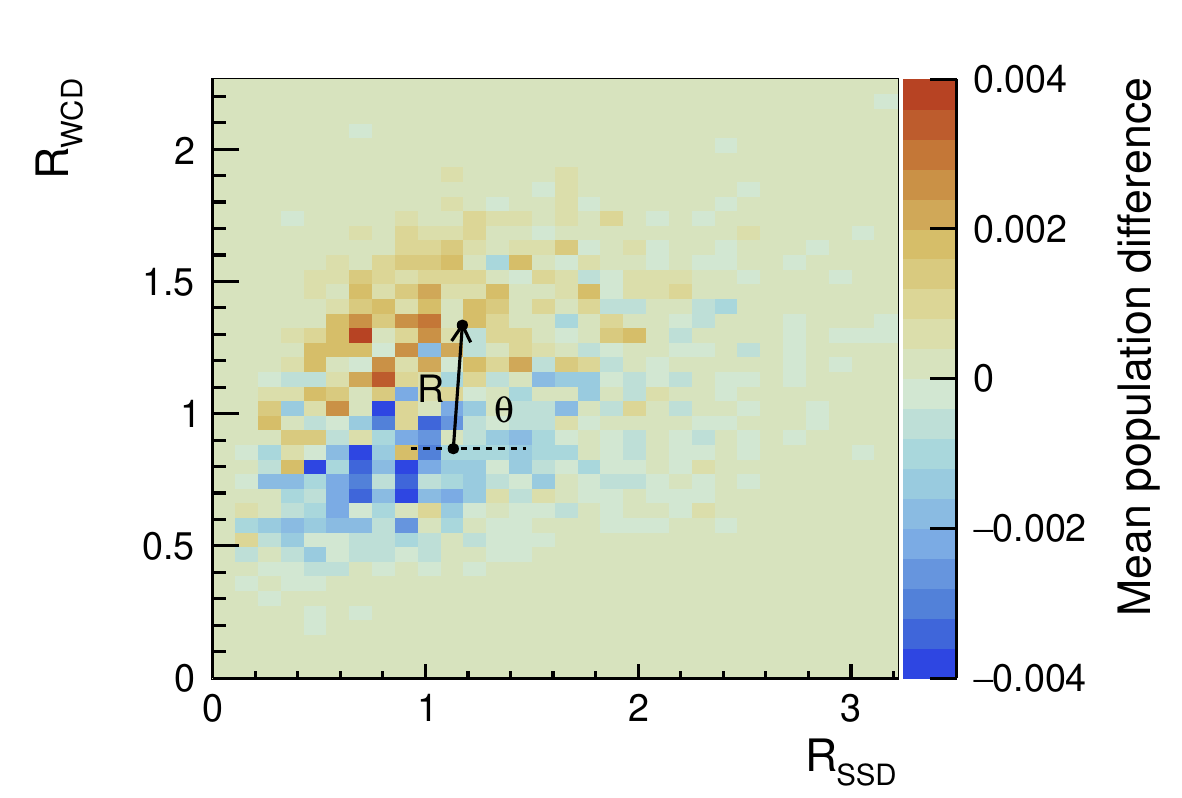}
 \includegraphics[width=0.49\linewidth]{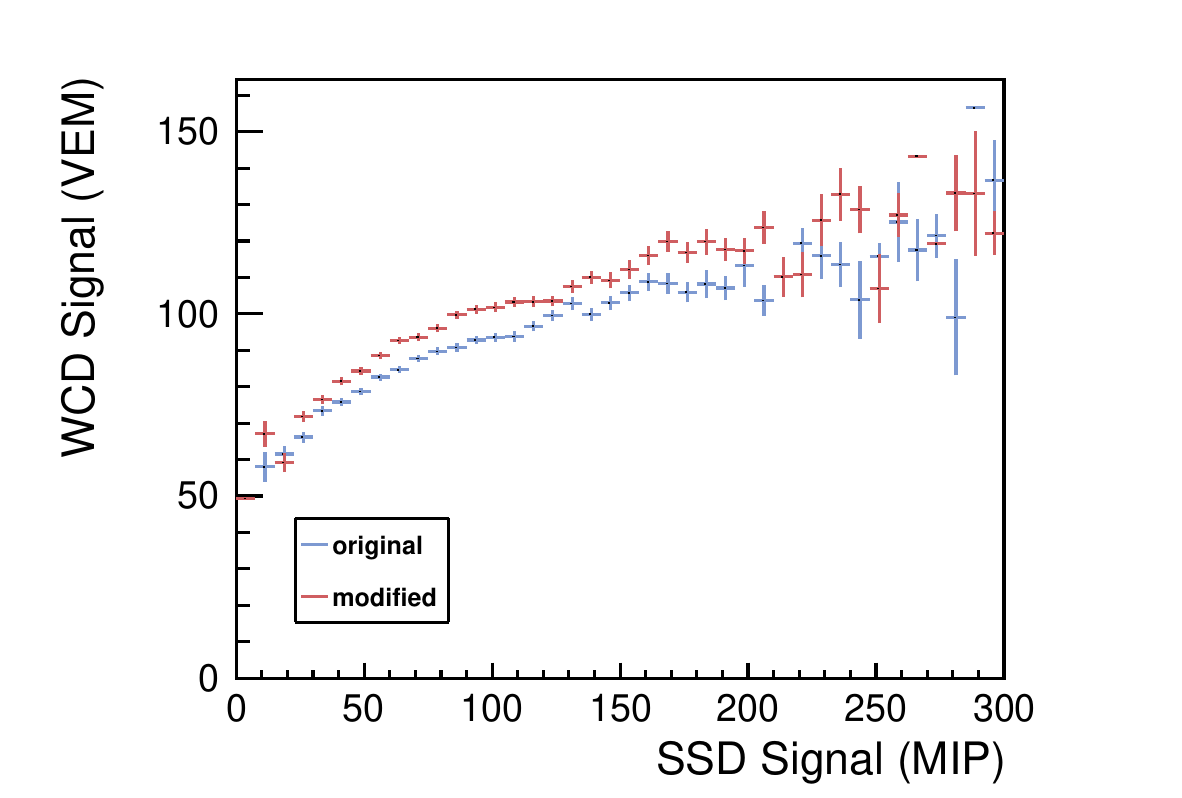}
 \caption{\label{fig:emSSDvsWCD} Modification of the high-energy tail of the electromagnetic: Difference in the SSD-WCD calibration plot between the modified and original distributions with a scheme illustrating the geometric representation of R and $\theta$ (left). $ R_{SSD}$ and $ R_{WCD}$ denote the detector signals normalized to the mean signal of the corresponding original distribution.; WCD signal profile as a function of the SSD signal (right).}
\end{figure}

Figure \ref{fig:emSSDvsWCD} (left) presents a difference plot illustrating the impact of a specific modification (intensity $c=2$) on the correlation between SSD and WCD signals. This plot visually shows that events associated with the modified distribution shift towards higher WCD signal values than the original distribution. This behaviour is expected since the WCD, functioning as a calorimeter for electromagnetic particles, is sensitive to particle energies. The modification increases the abundance of high-energy particles, yielding large electromagnetic showers in the tank and, consequently, a higher number of Cherenkov photons. Conversely, the SSD signal remains largely unaffected, as the total number of particles remains unchanged by the modification. This behaviour is further supported by the profile plot in Figure \ref{fig:emSSDvsWCD} (right), which demonstrates higher average WCD signal values for the modified distribution. 

\begin{figure}[ht]
 \centering
\includegraphics[width=0.49\linewidth]{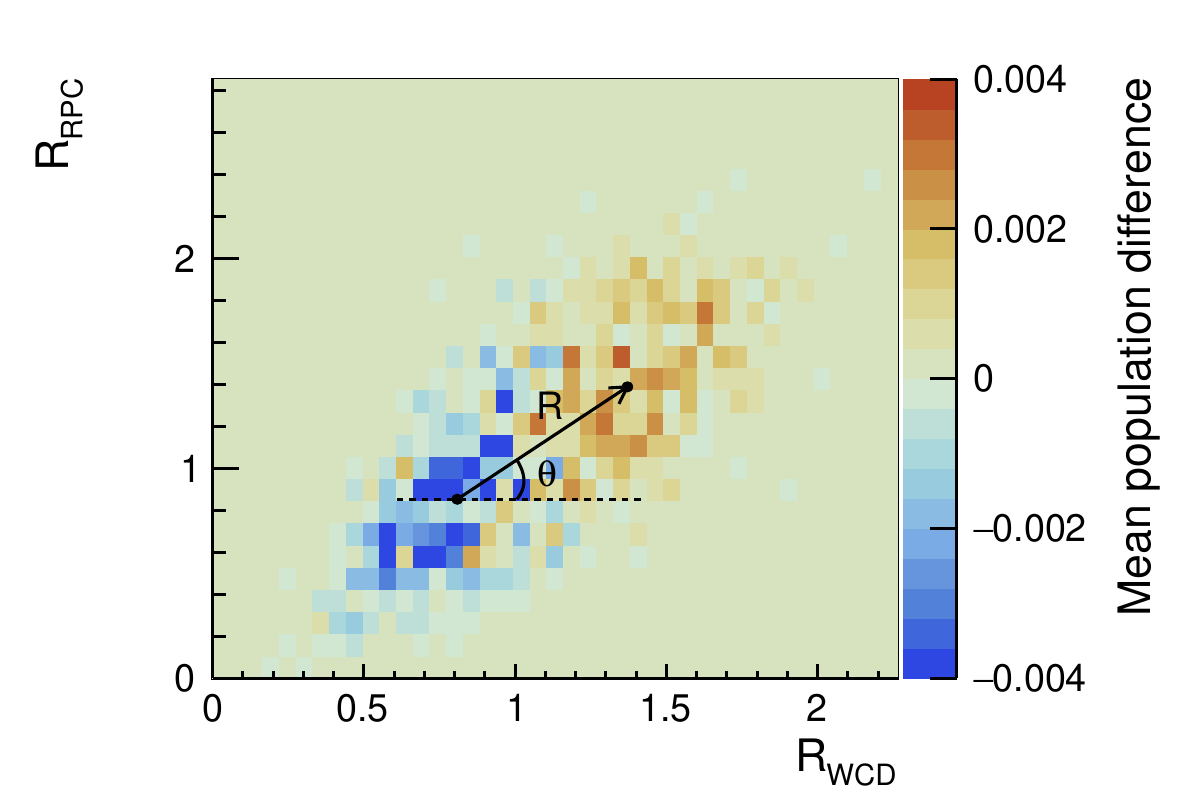}
\includegraphics[width=0.49\linewidth]{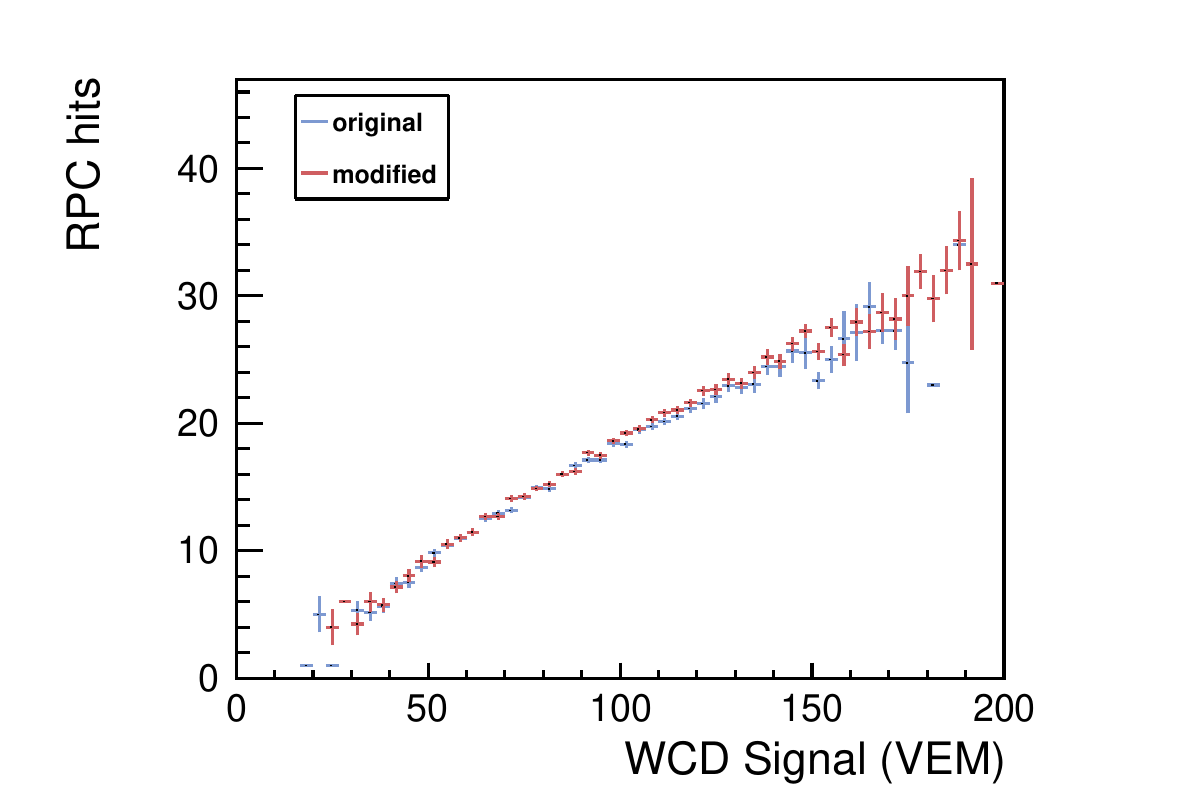}
 \caption{\label{fig:emWCDvsRPC} Modification of the high-energy e.m. spectrum tail:  Difference in the WCD-RPC calibration plot between the modified and original distributions with a scheme illustrating the geometric representation of R and $\theta$ (left); $ R_{WCD}$ and $ R_{RPC}$ denote the detector signals normalized to the mean signal of the corresponding original distribution. RPC hits profile as a function of the WCD signal (right).}
\end{figure}

Figure \ref{fig:emWCDvsRPC} (left) presents the detector response upon the above energy spectrum modification for the WCD and the RPCs. Contrary to the previous case, both WCD signals and RPC hits exhibit a consistent scaling with the modification, which is evident from the correlated displacement of the populations along both axes . This behaviour is expected for RPCs, as higher-energy particles generate larger showers that are more likely to penetrate the entire WCD tank and reach the RPC detector. The corresponding profile plot in Figure \ref{fig:emWCDvsRPC}  (right) illustrates the correlation between the outputs of the two detectors with this specific modification.

\begin{figure}[ht]
 \centering
 \includegraphics[width=0.8\textwidth]{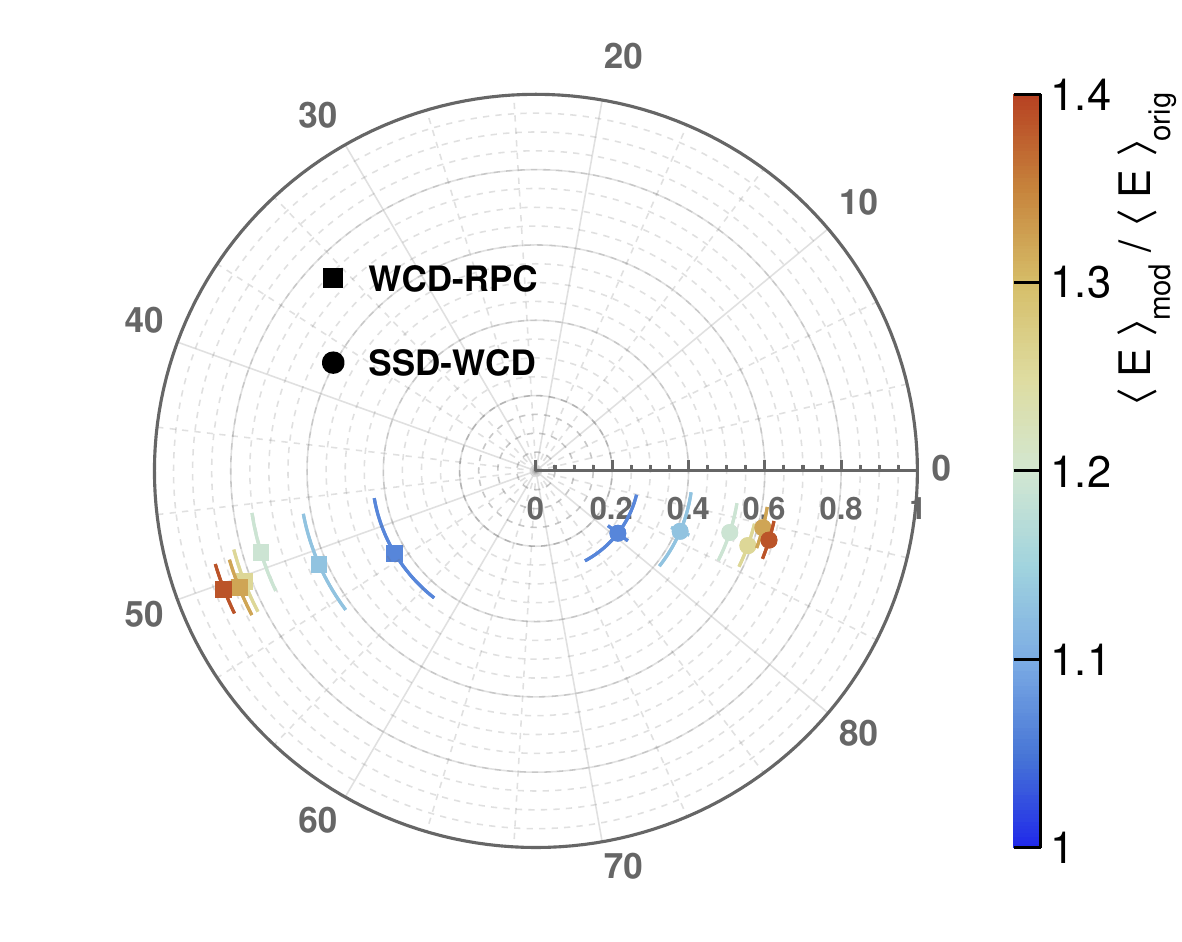}
 \caption{\label{fig:PolarEMhigh} Summary of the evolution of the  WCD-SSD and  RPC-WCD observables with the modification of the high-energy electromagnetic energy spectrum tail. $R$ is displayed along the radial axis while $\theta$ can be read in the angle axis.}
\end{figure}

To quantify the changes observed in Figs.~\ref{fig:emSSDvsWCD} and~\ref{fig:emWCDvsRPC}, we consider a series of modifications of the same type, each with varying intensities, i.e., different values of the parameter $c$. For each case, we calculate the estimators $R$ and $\theta$ defined in the previous section and present them in Fig.~\ref{fig:PolarEMhigh}. As seen in this figure, as the modification intensity increases, the distance between the barycenters of the points increases, resulting in an increase in $R$. Conversely, $\theta$ should serve as a distinctive indicator of the type of modification applied (see section \ref{subsection:caveats}) and the specific detectors under analysis.

Figure \ref{fig:PolarEMhigh} presents the results for a range of modifications with $c$ values from 1.5 to 6. To better understand these modifications, we categorized them based on the energy fraction between the modified and original distributions, a more physically intuitive variable.  The plot clearly illustrates the dependence of the $R$ variable on these modifications to the energy spectrum. In the SSD-WCD difference plots, $\theta$ is approximately $90^\circ$, indicating that the SSD signal remains largely unaffected by the modification. In contrast, for the WCD-RPC difference plots, $\theta$ is well-defined, reflecting the correlation between the scaling behaviour of these two detectors in response to the modification.

The results obtained here highlight the sensitivity of the detectors at this station to modifications in the energy distribution of electromagnetic particles. Moreover, this sensitivity can be quantified systematically and linked directly to the modification applied. The observed detector responses align with our prior understanding of their individual characteristics. 

\subsection{Low-energy tail of the muonic component}
\label{subsection:MuCompLE}

Similarly to the electromagnetic shower component, the muon energy spectrum exhibits two power-law tails—one at low energies and another at high energies—resulting in a roughly symmetrical distribution (see Fig.~\ref{fig:spectra} (right)). Notably, air shower muons have significantly higher average energies than electromagnetic particles. As a result, a typical air shower muon is relativistic and highly likely to traverse the entire WCD and reach the RPC. Considering these factors, we adopt a different approach to modifying the muon energy distribution. By increasing the relative weight of low-energy muons while maintaining a constant total particle count, we anticipate a corresponding decrease in WCD signals and RPC hits.

In this case, to ensure detector output is primarily dominated by muons, the station is fixed at a core distance of $r_{\rm core} = 700\,$ m. This larger distance from the shower core results in a reduced particle flux reaching the detectors, potentially impacting the statistical significance of the analysis. For this configuration, the region of the spectrum to be modified is defined by the parameters $a = 2.5\,$GeV and $b = 50\,$MeV. Figure \ref{fig:spectra} (right) illustrates an example of this type of modification with a magnitude of $c = 2$.

\begin{figure}[ht]
 \centering
\includegraphics[width=0.49\linewidth]{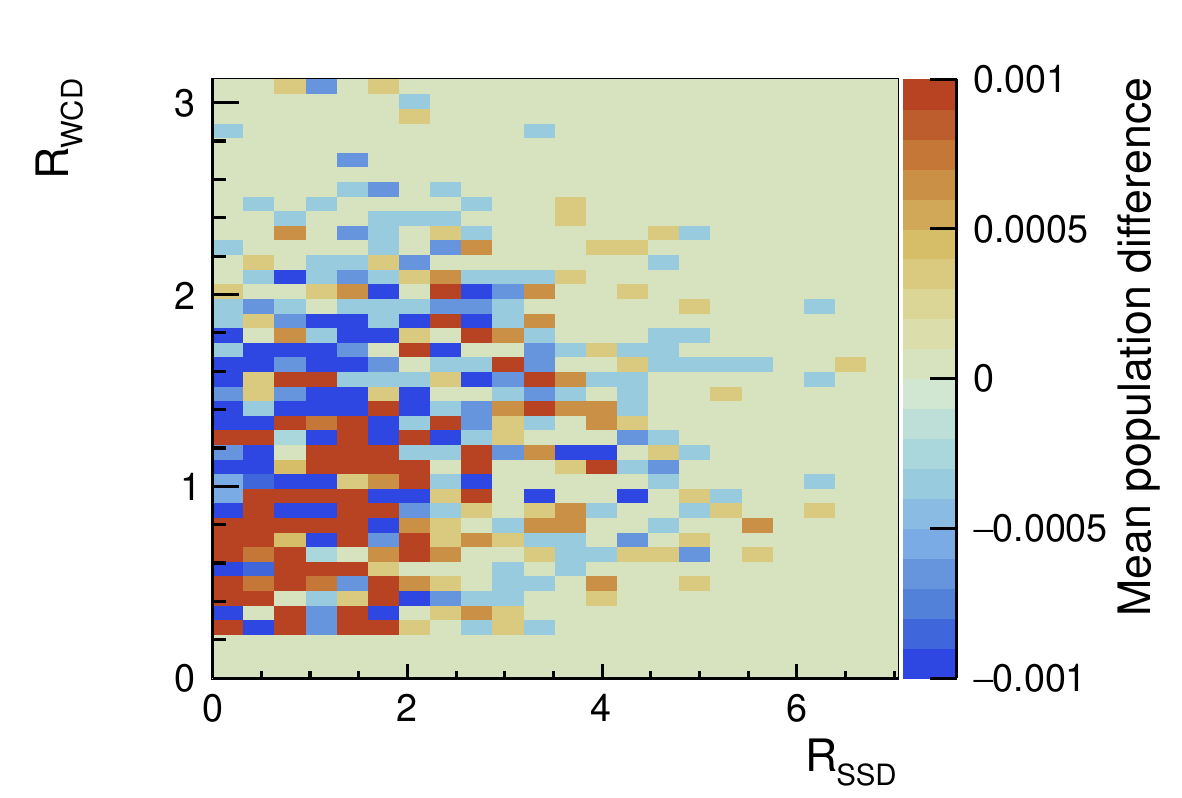}
\includegraphics[width=0.49\linewidth]{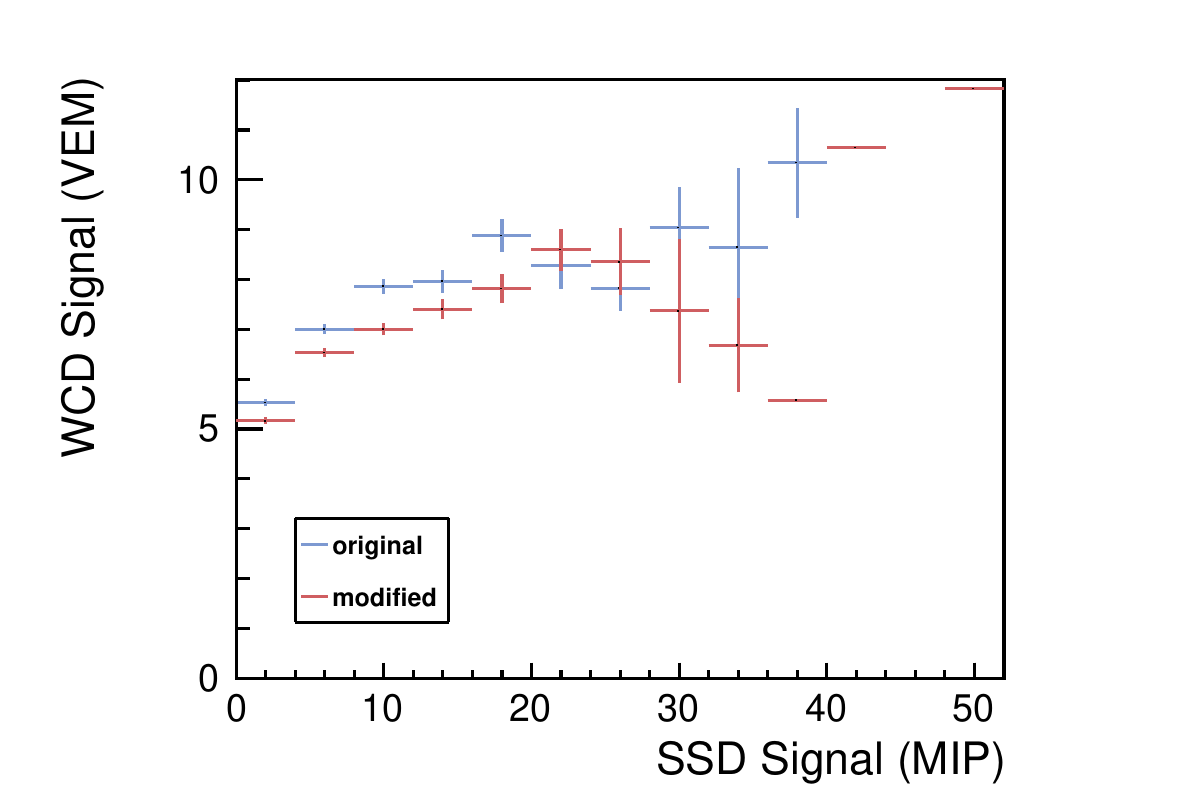}
 \caption{\label{fig:muSSDvsWCD} Modification of the low-energy tail of the muonic spectrum: Difference in the SSD-WCD calibration plot between the modified and original distributions (left). $ R_{SSD}$ and $ R_{WCD}$ denote the detector signals normalized to the mean signal of the corresponding original distribution; WCD signal profile as a function of the SSD signal(right) . }
\end{figure}

Figure~\ref{fig:muSSDvsWCD} (left) presents the difference plot corresponding to the detector response between SSD and WCD signals. The absence of events with a WCD signal below approximately $2\,$VEM is due to the applied station trigger. Unlike the modification to the high-energy tail of the electromagnetic particle energy distribution, this modification causes a downward shift in the WCD signal, consistent with our earlier discussion. As expected, the SSD signal remains largely insensitive to this type of modification. The effect of the modification is less apparent in the profile plot on the right of Figure  \ref{fig:muSSDvsWCD} (right) and is likely due to the truncation of the distribution at zero, which can influence the average values. Despite this, a clear visual separation persists between bins with positive and negative values in the difference plot, enabling the calculation of the estimators.  

\begin{figure}[ht]
 \centering
\includegraphics[width=0.49\linewidth]{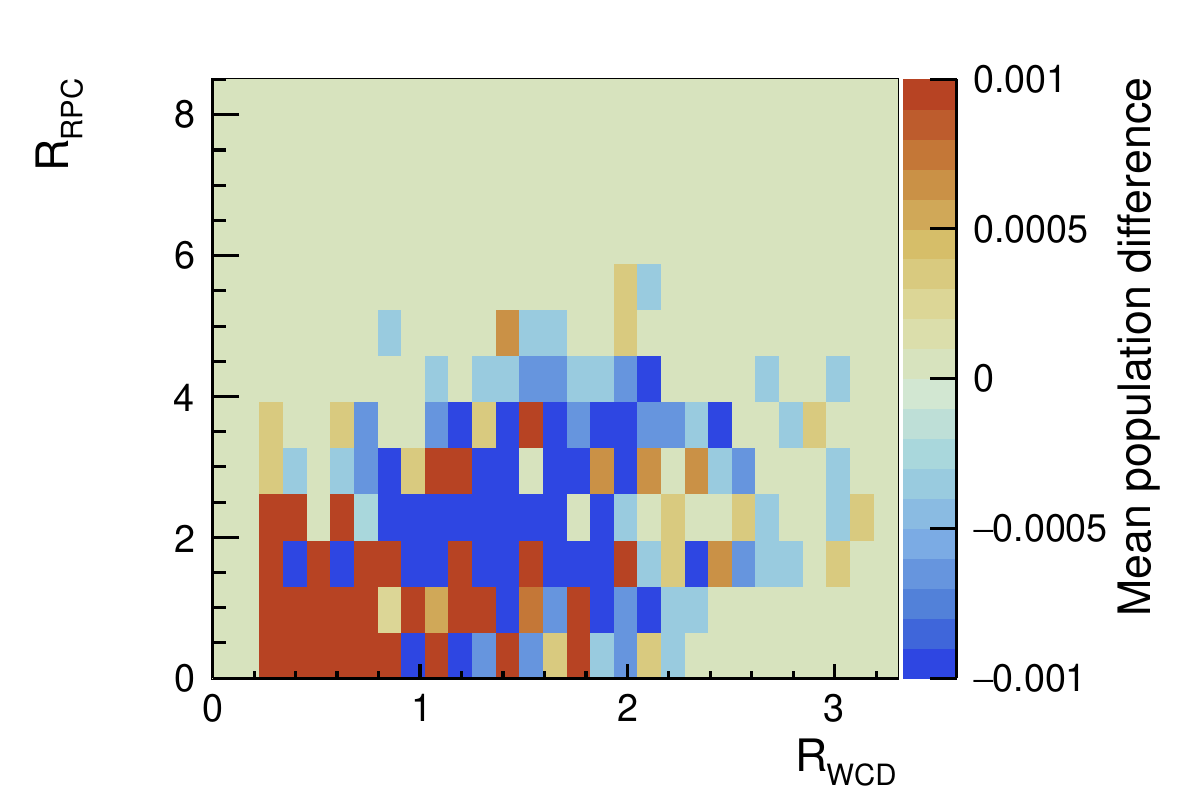}
\includegraphics[width=0.49\linewidth]{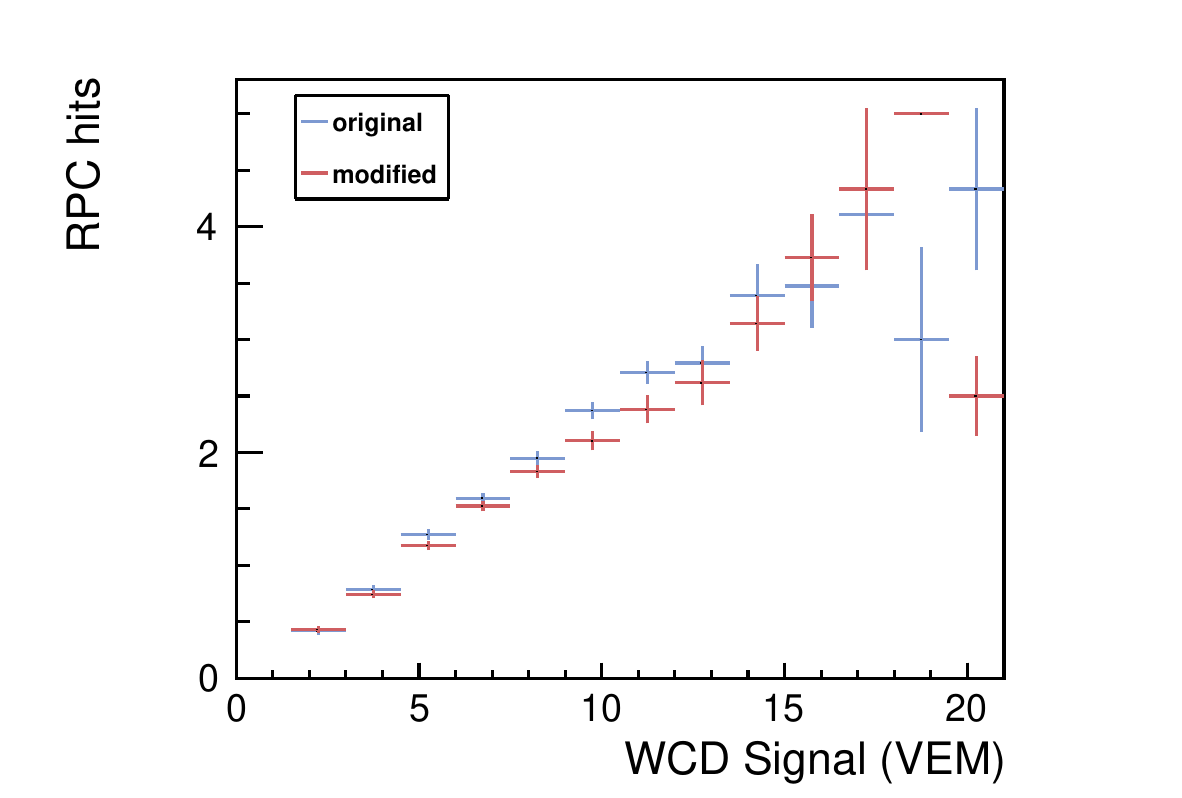}
 \caption{\label{muWCDvsRPC} Modification of the low-energy muonic spectrum tail: Difference in the WCD-RPC calibration plot between the modified and original distributions (left). $ R_{WCD}$ and $ R_{RPC}$ denote the detector signals normalized to the mean signal of the corresponding original distribution; RPC hits profile as a function of the WCD signal  (right) .}
\end{figure}

The corresponding difference plot for the WCD and RPC detector response is shown in Figure \ref{muWCDvsRPC} (left). Similar to the SSD-WCD difference plot, this one exhibits a downward shift along both axes, indicating that both detectors respond proportionally to the modification. This behaviour is expected, as the modified distribution contains relatively fewer relativistic muons capable of fully traversing the WCD and reaching the RPC. The profile plot in Fig.~\ref{muWCDvsRPC} (right) further highlights the correlation between the scaling of the two detectors under this modification.

\begin{figure}[ht]
 \centering
 \includegraphics[width=0.8\textwidth]{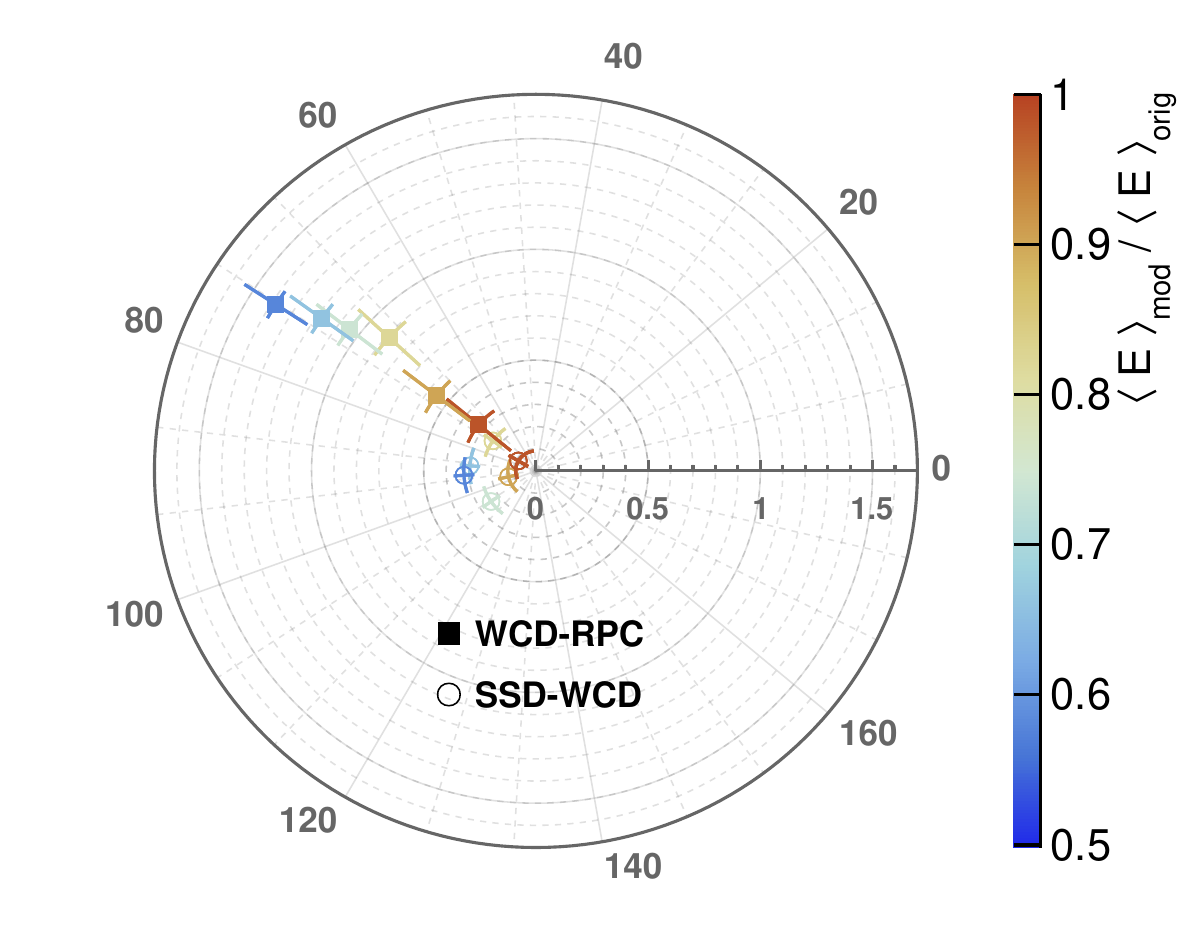}
 \caption{\label{fig:PolarMuLow} Summary of the evolution of the  WCD-SSD and  RPC-WCD observables with the modification of the low-energy tail of the muonic energy spectrum.}
\end{figure}

Following the approach used for modifying the high-energy tail of the electromagnetic particle energy distribution (Section \ref{subsec:emHEresults}), we considered a range of modifications with magnitudes from $c=1.5$ to $c=6$. For each case, the estimators $R$ and $\theta$ were calculated for both SSD-WCD and WCD-RPC difference plots. The results are presented in Figure \ref{fig:PolarMuLow}. Note that in this and all other radial plots shown in the article, the values of $\theta$ always lie within the principal angular range indicated by the plot axes. 

In this case, the WCD-RPC difference plots demonstrate a much clearer separation of $R$ values across different modification intensities when compared with the SSD-WCD difference plots. This effect is due to the RPC detector exhibiting significantly higher sensitivity to changes in the muonic component compared to the WCD, while the SSD remains largely insensitive to muonic variations. This sensitivity difference comes from the fact that at this distance from the shower core, the electromagnetic and muonic contributions to the WCD signal remain comparable, whereas the RPC signal is clearly dominated by muons, as shown in Figure 2.

Alongside this, for the SSD-WCD difference plot, the values of $\theta$ are less consistent, particularly at lower values of $c$. This is likely due to reduced statistical precision, a consequence of positioning the station farther from the shower core. 
Nevertheless, for the WCD-RPC difference plots, the values of $\theta$ remain consistent across all modification intensities, emphasizing the reliability of this detector combination. 

Overall, these results confirm the detector's sensitivity to modifications in the low-energy tail of the muon energy distribution, albeit to a lesser extent than for the electromagnetic component. Although the lower statistical significance introduces some uncertainty and the trigger imposes an abrupt event cut, the estimators remain robust under these conditions. 

\subsection{Overview}

Although the previous subsections focused on two specific modification configurations, numerous other scenarios were tested for both the electromagnetic and muonic energy distributions. These configurations primarily differ in two key aspects: the station's distance from the shower core and the spectral energy region being modified.

Table~\ref{TabTestsEM}  summarizes the results for all tested electromagnetic particle energy distribution modifications. As expected, the configuration with $r_{\rm core} = 320\,$m and modification to the high-energy tail yielded the highest detector sensitivity. At this distance, both the RPC and WCD detectors exhibit a clear response to the applied modifications. At larger distances to the shower core, the WCD remained sensitive, whereas the RPC did not. This is expected, as the MARTA RPC design is optimized for muon detection, and at greater distances, muons dominate over the electromagnetic component. As previously discussed, the SSD is essentially insensitive to these modifications due to the unchanged particle count.

Modifying the low-energy electromagnetic plateau presents significant challenges. Due to the sheer number of particles in this region, any alteration requires a substantial normalization factor, which inevitably impacts the high-energy tail. As a result, isolating modifications to this plateau without unintentionally altering the total particle count is extremely difficult. In Section \ref{sec:Exp}, we briefly explore the effects of such modifications without imposing a fixed particle count constraint.

\renewcommand{\arraystretch}{1.5}
\begin{table*}
  \begin{center}
    \caption{Summary of results for all modification configurations of the electromagnetic component.
Arrow direction ($\uparrow$/$\downarrow$) indicates whether the applied modification 
increased or decreased the detector signal. Coloured squares represent detector 
sensitivity levels: green denotes sensitivity (darker shades = stronger responses), 
while red indicates no observable sensitivity.}
    \vspace{0.5cm} 
    \begin{tabular}{lcccccccc}
      \hline\hline
      \multirow{2}{*}{\textbf{EM}} & \multicolumn{3}{c}{\textbf{High Energies}} & \multicolumn{3}{c}{\textbf{Low Energies}} \\ 
      & \textbf{WCD} & \textbf{SSD} & \textbf{RPC} &\textbf{WCD} & \textbf{SSD} & \textbf{RPC} & \\
      \hline
      \textbf{320 m} & $\uparrow$ \textcolor{ForestGreen!80}{$\blacksquare$} & -- \textcolor{Maroon!80}{$\blacksquare$} & $\uparrow$ \textcolor{ForestGreen!50}{$\blacksquare$} & -- \textcolor{Maroon!80}{$\blacksquare$} & -- \textcolor{Maroon!80}{$\blacksquare$} & -- \textcolor{Maroon!80}{$\blacksquare$} &  \\
      \textbf{700 m} & $\uparrow$ \textcolor{ForestGreen!50}{$\blacksquare$} & -- \textcolor{Maroon!80}{$\blacksquare$} & -- \textcolor{Maroon!80}{$\blacksquare$} & -- \textcolor{Maroon!80}{$\blacksquare$} & -- \textcolor{Maroon!80}{$\blacksquare$} & -- \textcolor{Maroon!80}{$\blacksquare$} & \\
      \textbf{1060 m} & $\uparrow$ \textcolor{ForestGreen!20}{$\blacksquare$} & -- \textcolor{Maroon!80}{$\blacksquare$} & -- \textcolor{Maroon!80}{$\blacksquare$} & -- \textcolor{Maroon!80}{$\blacksquare$} & -- \textcolor{Maroon!80}{$\blacksquare$} & -- \textcolor{Maroon!80}{$\blacksquare$} & \\
      \hline\hline
    \end{tabular}
    \label{TabTestsEM}
  \end{center}
\end{table*}

\renewcommand{\arraystretch}{1.5}
\begin{table*}
  \begin{center}
    \caption{Summary of results for all modification configurations of the muonic component. 
Arrow direction ($\uparrow$/$\downarrow$) indicates whether the applied modification 
increased or decreased the detector signal. Coloured squares represent detector 
sensitivity levels: green denotes sensitivity (darker shades = stronger responses), 
while red indicates no observable sensitivity.}
    \vspace{0.5cm} 
    \begin{tabular}{lcccccccc}
      \hline\hline
      \multirow{2}{*}{\textbf{Muon}} & \multicolumn{3}{c}{\textbf{High Energies}} & \multicolumn{3}{c}{\textbf{Low Energies}} \\ 
      & \textbf{WCD} & \textbf{SSD} & \textbf{RPC} &\textbf{WCD} & \textbf{SSD} & \textbf{RPC} & \\
      \hline
      \textbf{320 m} & -- \textcolor{Maroon!80}{$\blacksquare$} &  -- \textcolor{Maroon!80}{$\blacksquare$} & -- \textcolor{Maroon!80}{$\blacksquare$} & -- \textcolor{Maroon!80}{$\blacksquare$} & -- \textcolor{Maroon!80}{$\blacksquare$} & -- \textcolor{Maroon!80}{$\blacksquare$} &  \\
      \textbf{700 m} & -- \textcolor{Maroon!80}{$\blacksquare$} & -- \textcolor{Maroon!80}{$\blacksquare$} & -- \textcolor{Maroon!80}{$\blacksquare$} & $\downarrow$ \textcolor{ForestGreen!20}{$\blacksquare$} & -- \textcolor{Maroon!80}{$\blacksquare$} & $\downarrow$ \textcolor{ForestGreen!20}{$\blacksquare$} & \\
      \textbf{1060 m} & -- \textcolor{Maroon!80}{$\blacksquare$} & -- \textcolor{Maroon!80}{$\blacksquare$} & -- \textcolor{Maroon!80}{$\blacksquare$} & $\downarrow$ \textcolor{ForestGreen!20}{$\blacksquare$} & -- \textcolor{Maroon!80}{$\blacksquare$} & $\downarrow$  \textcolor{ForestGreen!20}{$\blacksquare$} & \\
      \hline\hline
    \end{tabular}
    \label{TabTestsMu}
  \end{center}
\end{table*}

The results of the modifications to the muon energy distribution across all tested configurations are summarized in Table \ref{TabTestsMu}. As discussed in Section \ref{subsection:MuCompLE}, significant detector sensitivity is observed primarily for modifications to the low-energy tail of the muon energy spectrum and when the station is located far from the shower core, where the electromagnetic component is attenuated. Under these conditions, both the WCD and RPC detectors exhibit sensitivity, although to a lesser extent than in the case of electromagnetic modifications (as indicated by the lighter shade of green in the table). This reduced sensitivity is likely due to the lower number of particles reaching the detector at greater distances from the shower core.

Finally, the results from Tables \ref{TabTestsEM} and \ref{TabTestsMu} confirm that the two configurations highlighted in Sections \ref{subsec:emHEresults} (for the electromagnetic component) and \ref{subsection:MuCompLE} (for the muonic component) yield the strongest detector responses. These configurations were selected as they provide optimal conditions for studying each component individually. Notably, this also suggests that by applying different \( r_{\rm core} \) cuts, we can effectively isolate and analyze the distinct components of the shower.

\section{Experimental considerations}
\label{sec:Exp}

The results presented in Section~\ref{sec:Results} were obtained under \emph{ideal} conditions, meaning no reconstruction effects were considered. However, in realistic experimental scenarios, various factors may contribute to the degradation of these results. In this section, we evaluate the key factors that could impact the proposed measurement. 
Given that the experimental factors under investigation should remain independent of the chosen configuration, we decided to conduct these tests using the configuration described in Section~\ref{subsec:emHEresults}, i.e., for modifications on the high-energy tail of the electromagnetic energy spectrum.

\subsection{Energy reconstruction}

The analysis presented in Section~\ref{subsec:emHEresults} was repeated using an energy bin of $\log(E/{\rm eV}) \in [17.4, 17.6]$ to assess the impact of energy reconstruction on the measurement of the R and $\theta$ variables as opposed to a fixed energy of$\log(E/{\rm eV}) = 17.5$ The chosen width of the energy bin aims to replicate the typical energy resolution achieved at these scales~\cite{PierreAuger:2021hun}.

As illustrated in Fig.~\ref{fig:emRecPolar} (top, left), although there is a slight shift in the data points, especially for $R$, the sensitivity to the energy spectrum represented by changes to the $c$-parameter remains substantially unchanged.

\subsection{Shower geometry dependence}
\label{subsec:core}
The dependence on shower inclination has been investigated by analyzing showers with zenith angles of $30^\circ$ and $50^\circ$. The first value was chosen as it represents the mean expected angle for showers recorded by this station, while the latter was selected to be sufficiently different yet still physically realistic. It is important to note that, for this method to be effective, the majority of shower particles must traverse all three detectors. Additionally, since both the SSD and RPC are planar detectors, their effective area decreases rapidly with increasing shower inclination.

The obtained results are presented in Fig.~\ref{fig:emRecPolar} (top, right). As expected, noticeable changes in the $R$ and $\theta$ parameters are observed, yet the sensitivity to the energy spectrum remains. Given that the zenith angle resolution achieved in these arrays at these energies is of $\mathcal{O}(1^\circ)$ and that the evolution with the zenith angle appears smooth, we anticipate that the experimental uncertainty on the angular resolution should not preclude such a measurement.

\subsection{Reconstruction of the shower core position}

\begin{figure}[ht]
 \centering
\includegraphics[width=0.45\linewidth]{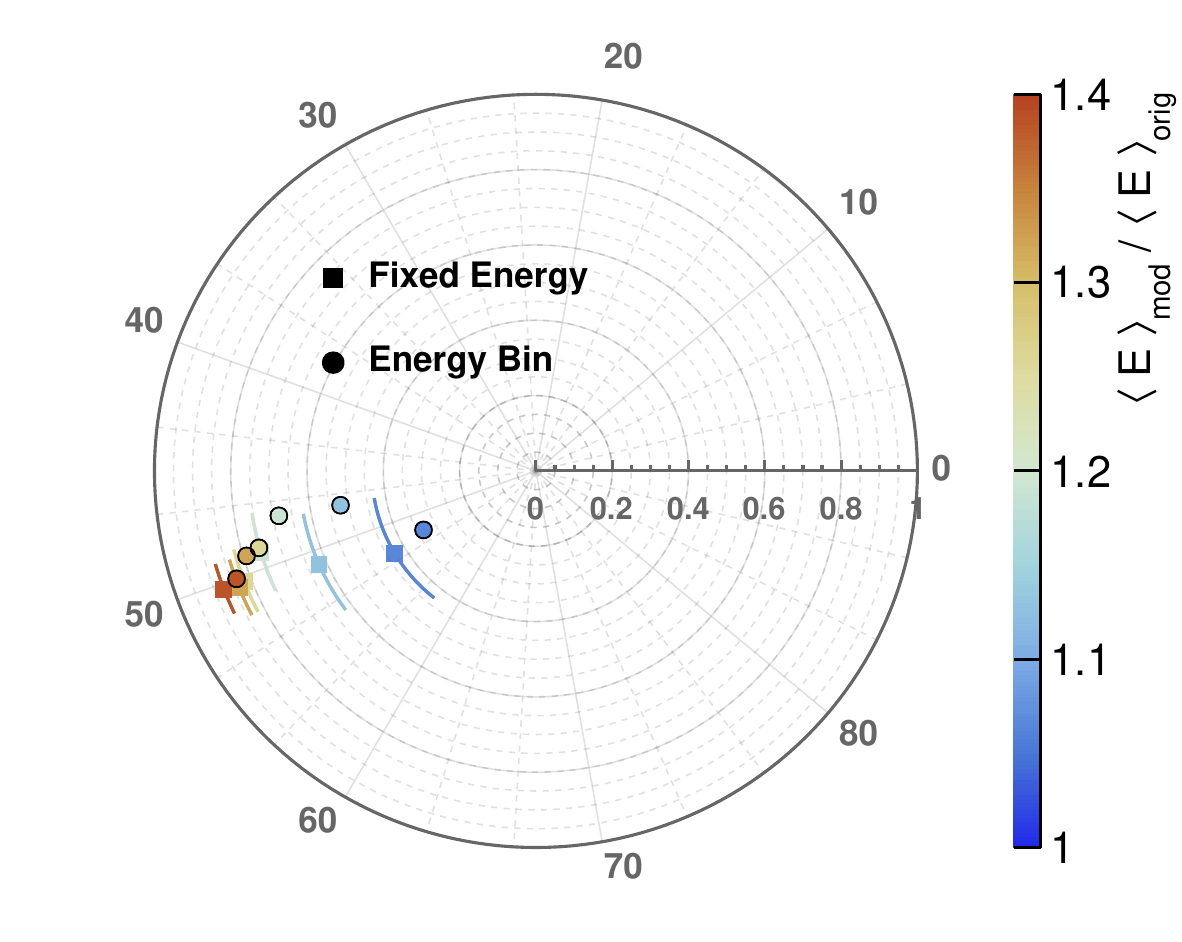}
\includegraphics[width=0.45\linewidth]{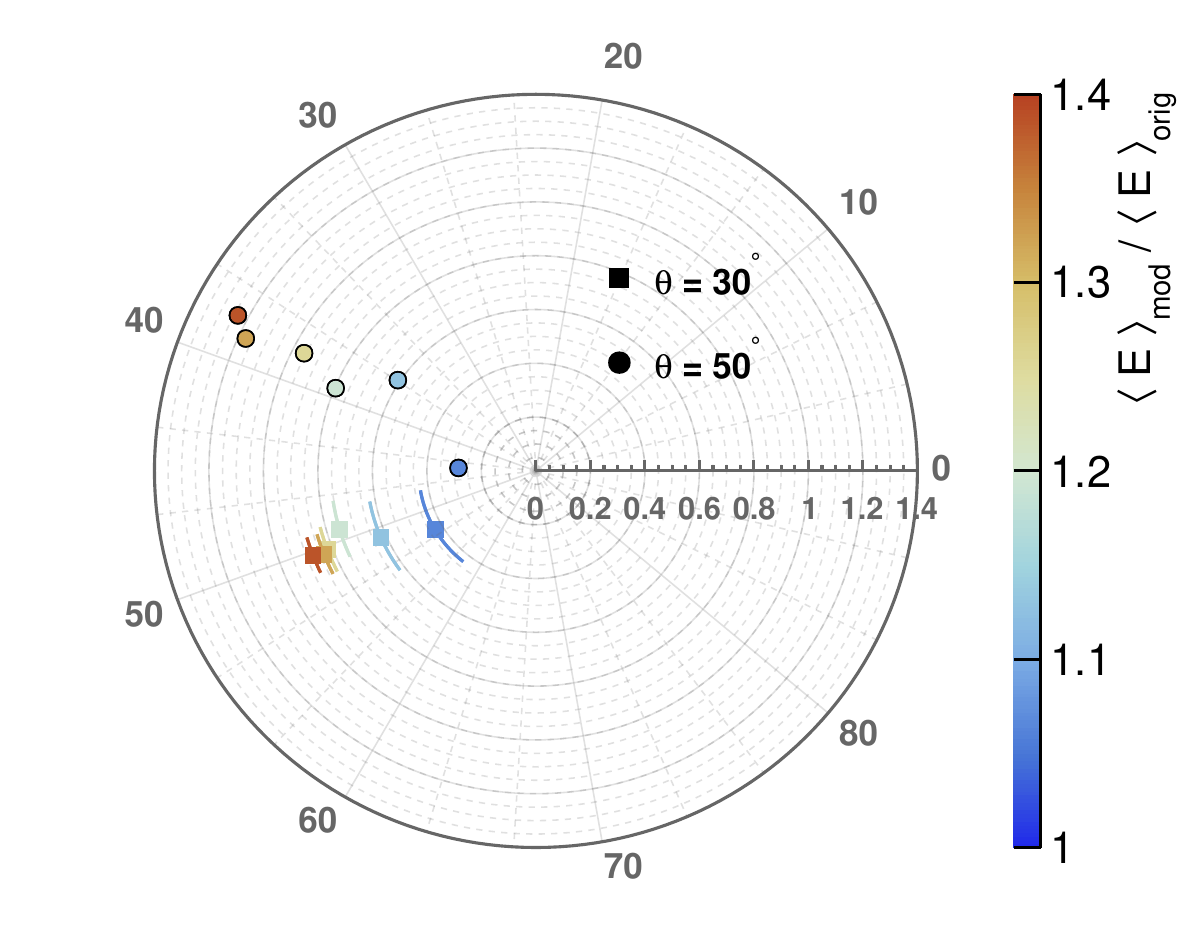}
\includegraphics[width=0.45\linewidth]{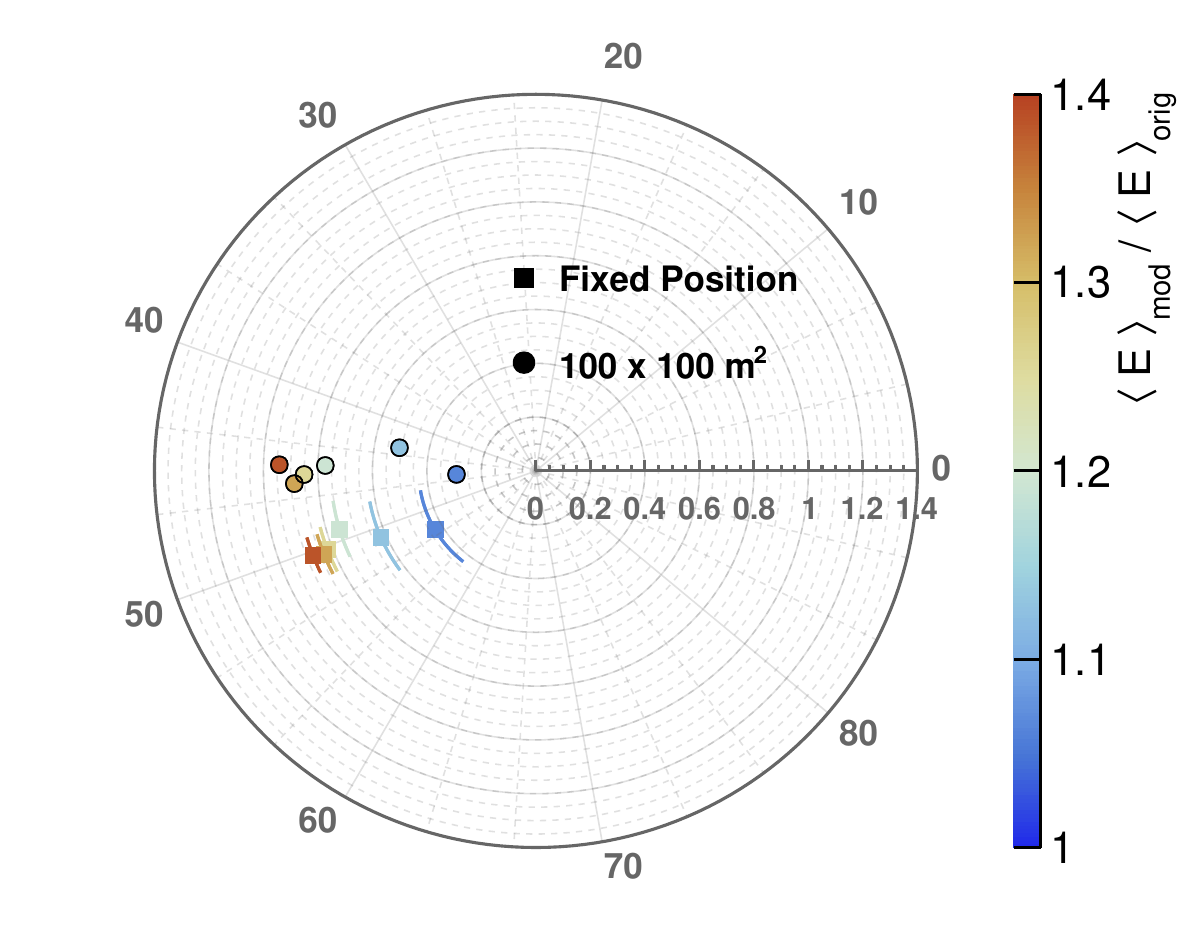}
 \includegraphics[width=0.45\linewidth]{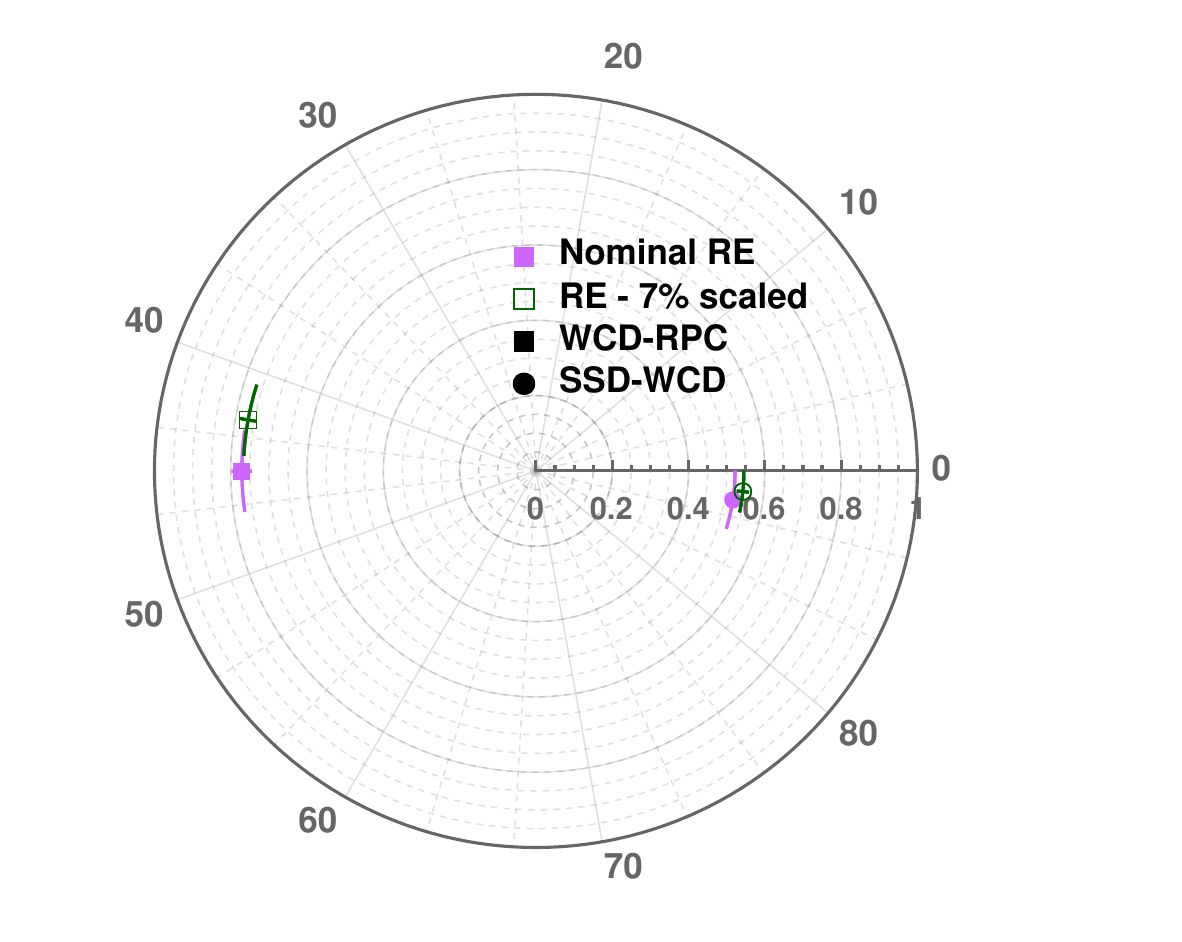}
 \caption{\label{fig:emRecPolar} Summary of the dependence of the observables on shower energy reconstruction resolution (top left), shower direction dependence (top right), shower core reconstruction resolution (bottom left), and changes in the reflectivity of the WCD Tyvek (bottom right) (see caption and text for details). The observables considered here are derived from a WCD-SSD difference plot corresponding to a modification in the high-energy tail of the electromagnetic particle energy distribution.}
\end{figure}

Finally, the last reconstruction parameter tested was the shower core position. Based on~\cite{PierreAuger:2015eyc}, we expect the resolution to be of the order of $\sigma_c \sim 50\,{\rm m}$. As a highly conservative test, we uniformly sampled the shower core distribution over a $100 \times 100\,{\rm m^2}$ area, leading to the results presented in Fig.~\ref{fig:emRecPolar} (bottom, left). Once again, no significant scrambling of the $R$ and $\theta$ variables is observed that would hinder the assessment of the electromagnetic shower particle energy spectrum.

\subsection{Caveats on the analysis}
\label{subsection:caveats}

The analysis presented in this paper assumes that the multiplicity of shower particles at the ground is constant and that the detector calibrations are perfectly known. These are strong assumptions, given the existing uncertainties associated with WCD~\cite{PierreAuger:2020flu} and the uncertainties in shower physics mechanisms~\cite{whispMuonDeficit}.  

The light collection in the WCD depends on optical parameters such as the photosensor light collection efficiency, the Cherenkov light attenuation in water, and the Tyvek reflectivity. These quantities are typically calibrated by leveraging the constant omnidirectional atmospheric muon flux and correlating the measured signal peak with the signal from vertically crossing muons, commonly called Vertical Equivalent Muons (VEM)~\cite{PierreAuger:2005znw}. While this procedure allows for calibration of the mean signal, long-term observations have shown that WCD optical parameters undergo ageing, as evidenced by variations in the photosensor signal time-trace peak intensity relative to its charge~\cite{PierreAuger:2020flu}. One possible explanation for this effect is the degradation of Tyvek reflectivity.

We artificially degraded the reflectivity by approximately $7\%$ to emulate this effect and evaluated its impact on the $R$ and $\theta$ parameters after applying the VEM calibration procedure (see Fig.~\ref{fig:emRecPolar}, bottom right). The observed changes compared to the nominal Tyvek reflectivity were negligible. Moreover, it is important to note that any modifications in the energy spectrum should manifest consistently across the three-detector system, further reinforcing the confidence that any observed variations stem from changes in shower behaviour rather than detector calibration issues.

\begin{figure}[ht]
 \centering
\includegraphics[width=0.45\linewidth]{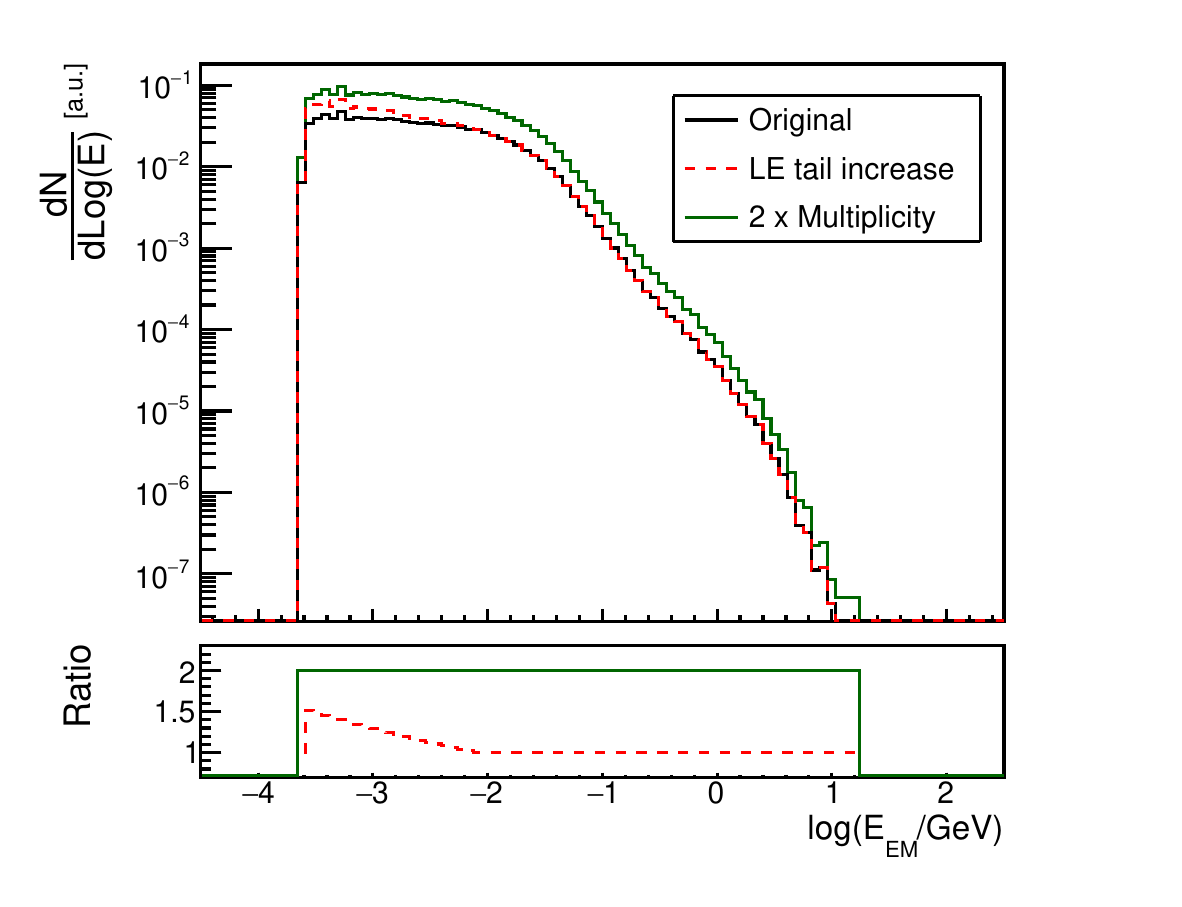}
\includegraphics[width=0.45\linewidth]{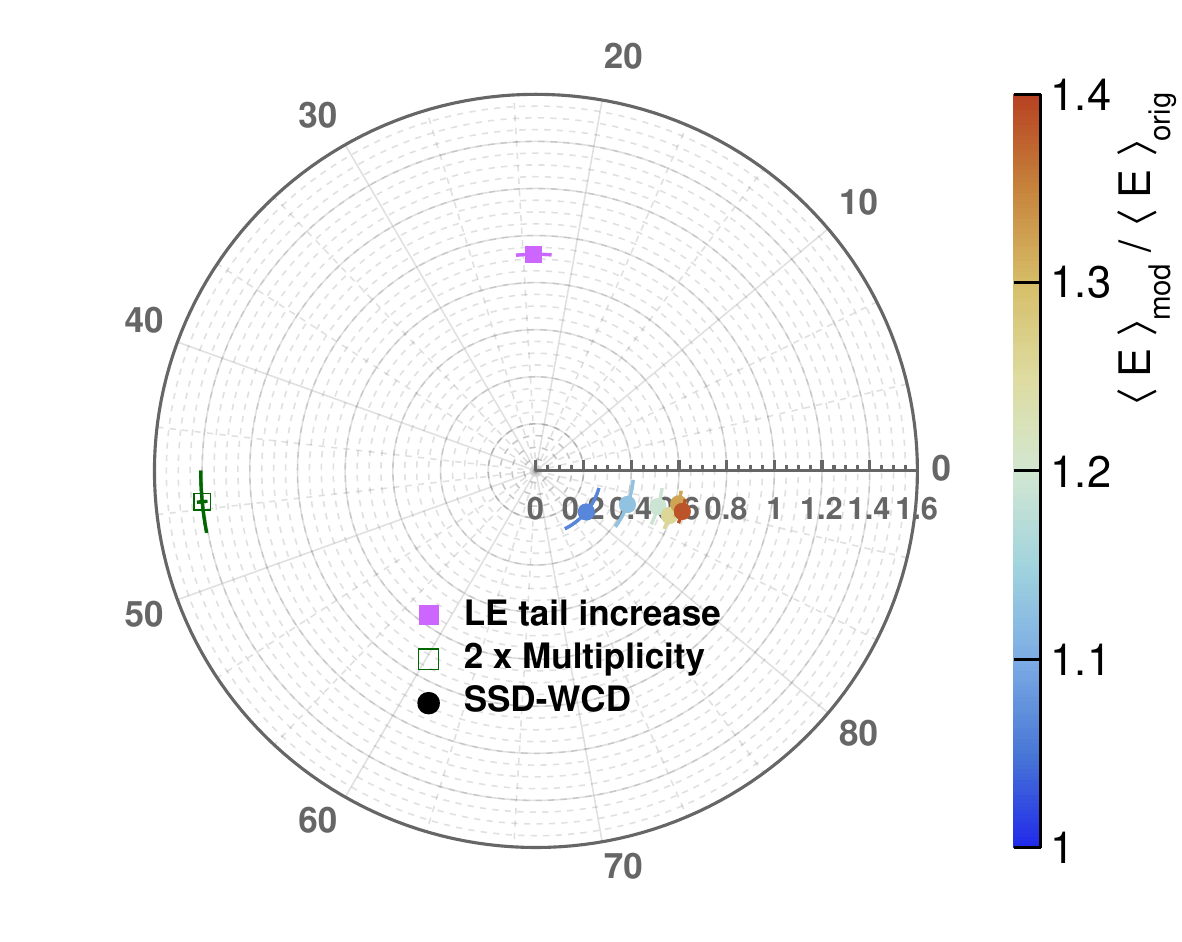}
 \caption{\label{fig:emHighMultiplicity} Modifications to the electromagnetic particle energy distribution with no conservation of particle number (left). Summary of the dependence of WCD-SSD spectrum observables on shower multiplicity (see text for details).}
\end{figure}

Two scenarios were evaluated to assess the impact of the shower multiplicity on this method: one in which the overall multiplicity is increased by a factor of two, and another in which only the number of low-energy particles (LE tail) is increased (see Fig.~\ref{fig:emHighMultiplicity} (left)).  

The results, shown in Fig.~\ref{fig:emHighMultiplicity} (right), clearly indicate that while $R$ changes significantly, so does $\theta$, strongly signalling an increase in the number of particles. This increase in $\theta$ can be readily explained by the enhanced SSD signal, which, as previously mentioned, is proportional to the number of particles crossing the scintillator.

Fixing the multiplicity while allowing the energy spectrum to vary inherently leads to a local violation of energy conservation at the specific station. However, this does not necessarily imply a violation of the total shower energy, as the lateral distribution function may deviate from the expected shape. This approach effectively examines whether the energy spectrum at a fixed distance from the shower core differs from simulation predictions.

Finally, this study was conducted using a WCD surface array with stations spaced $750\,$m apart, where a single station is equipped with an SSD on top and RPCs beneath the WCD. For this array configuration and the explored energies of $\log(E/{\rm eV}) = 17.5$, the expected event collection rate is approximately $1000$ events per month.

The statistical power of the \( R \) variable in distinguishing between different shower secondary energy spectra, characterized by varying \( c \)-parameters, was assessed. The results, as discussed in Sec.~\ref{subsec:emHEresults}, indicate that a sample of 1000 shower events is sufficient for this purpose.  

Given that the analysis can be conducted in bins of station distance to the shower core, with a resolution of approximately $100\,$m (see Sec.~\ref{subsec:core}), this result suggests that an estimation of the high-energy tail of the electromagnetic shower component and the low-energy tail of the muonic shower component could be achieved within roughly one year of data collection.

\section{Conclusions}
\label{sec:Conclusions}

In this study, we introduced a novel approach for assessing the energy spectrum of extensive air showers (EAS) components using a multi-hybrid detector station. By combining measurements from a scintillator surface detector (SSD), a water-Cherenkov detector (WCD), and resistive plate chambers (RPCs), we demonstrated the feasibility of disentangling the electromagnetic and muonic components of air showers at ground level.

Our analysis showed that the response of the WCD correlates strongly with particle energy, while the SSD remains relatively insensitive to energy variations and primarily acts as a particle counter. The additional inclusion of the RPCs provided an independent confirmation of the muonic component. By leveraging these complementary responses, we established a methodology to infer the energy distribution of secondary particles in air showers.

The high-energy tail of the electromagnetic spectrum can be successfully characterized by examining variations in the SSD-WCD correlation. We found that an increase in high-energy particles leads to a clear shift in the WCD response, while the SSD signal remains largely unaffected. Similarly, the low-energy tail of the muon spectrum was studied using stations positioned farther from the shower core, where the muonic component dominates. A modification in the low-energy muon population resulted in a systematic change in both WCD and RPC signals, confirming the method’s sensitivity to this region of the spectrum.

We investigated the impact of energy resolution, shower geometry reconstruction, and core position uncertainties to assess potential experimental limitations. The results indicate that the proposed method is robust to these effects, with only minor variations in the extracted energy spectra. Additionally, we tested detector ageing effects, such as the degradation of Tyvek reflectivity in the WCD, and found that they do not significantly affect the extracted shower parameters.

Overall, this work demonstrates that a single multi-hybrid station, when embedded within a sparse surface array, can provide valuable insights into the energy spectrum of shower components. These findings open new avenues for future experimental efforts, particularly in the context of hybrid cosmic ray observatories. For instance, within the Pierre Auger Observatory Infill array, the Underground Muon Detector (UMD)~\cite{PierreAuger:2020gxz} could also contribute to a multi-hybrid configuration through its complementary sensitivity to muons. Its buried location suppresses the electromagnetic component more effectively and selects, on average, higher-energy muons than the water Cherenkov detectors, thus providing additional information. A dedicated study would, however, be required to properly assess its contribution, which is beyond the scope of the present work. Further studies with real data, along with refinements in the reconstruction algorithms, will be crucial in validating this methodology under actual observational conditions.

\acknowledgments

We would like to thank Kevin Almeida Cheminant, Carola Dobrigkeit, M\'ario Pimenta and Jakub V\'icha for carefully reading the manuscript and providing useful comments. We would also like to thank the Pierre Auger Collaboration for the helpful discussions.
This work has been funded by Fundação para a Ciência e Tecnologia, Portugal, under project \url{https://doi.org/10.54499/2024.06879.CERN}. M.F. acknowledges financial support from FCT through grant 2025.00691.BD.

\bibliographystyle{JHEP}
\bibliography{references.bib}

\providecommand{\href}[2]{#2}\begingroup\raggedright\begin{thebibliography}{10}

\bibitem{PierreAuger:2024neu}
{\scshape Pierre Auger} collaboration, \emph{{Testing hadronic-model predictions of depth of maximum of air-shower profiles and ground-particle signals using hybrid data of the Pierre Auger Observatory}}, \href{https://doi.org/10.1103/PhysRevD.109.102001}{\emph{Phys. Rev. D} {\bfseries 109} (2024) 102001} [\href{https://arxiv.org/abs/2401.10740}{{\ttfamily 2401.10740}}].

\bibitem{PierreAuger:2021qsd}
{\scshape Pierre Auger} collaboration, \emph{{Measurement of the Fluctuations in the Number of Muons in Extensive Air Showers with the Pierre Auger Observatory}}, \href{https://doi.org/10.1103/PhysRevLett.126.152002}{\emph{Phys. Rev. Lett.} {\bfseries 126} (2021) 152002} [\href{https://arxiv.org/abs/2102.07797}{{\ttfamily 2102.07797}}].

\bibitem{whispMuonDeficit}
J.C.~Arteaga~Velazquez, \emph{{A report by the WHISP working group on the combined analysis of muon data at cosmic-ray energies above 1 PeV}}, \href{https://doi.org/10.22323/1.444.0466}{\emph{PoS} {\bfseries ICRC2023} (2023) 466}.

\bibitem{Castellina:2019irv}
{\scshape Pierre Auger} collaboration, \emph{{AugerPrime: the Pierre Auger Observatory Upgrade}}, \href{https://doi.org/10.1051/epjconf/201921006002}{\emph{EPJ Web Conf.} {\bfseries 210} (2019) 06002} [\href{https://arxiv.org/abs/1905.04472}{{\ttfamily 1905.04472}}].

\bibitem{pO_coll_LHC}
O.~Adriani, E.~Berti, L.~Bonechi, M.~Bongi, R.~D'Alessandro, G.~Castellini et~al., \emph{{LHCf plan for proton-oxygen collisions at LHC}}, \href{https://doi.org/10.22323/1.395.0348}{\emph{PoS} {\bfseries ICRC2021} (2021) 348}.

\bibitem{EM_all_2}
S.~Lafebre, R.~Engel, H.~Falcke, J.~Horandel, T.~Huege, J.~Kuijpers et~al., \emph{Universality of electron–positron distributions in extensive air showers}, \href{https://doi.org/10.1016/j.astropartphys.2009.02.002}{\emph{Astroparticle Physics} {\bfseries 31} (2009) 243}.

\bibitem{Śmiałkowski_2018}
A.~Śmiałkowski and M.~Giller, \emph{Universality of electron distributions in extensive air showers}, \href{https://doi.org/10.3847/1538-4357/aaa488}{\emph{The Astrophysical Journal} {\bfseries 854} (2018) 48}.

\bibitem{GILLER201592}
M.~Giller, A.~Śmiałkowski and G.~Wieczorek, \emph{An extended universality of electron distributions in cosmic ray showers of high energies and its application}, \href{https://doi.org/https://doi.org/10.1016/j.astropartphys.2014.04.003}{\emph{Astroparticle Physics} {\bfseries 60} (2015) 92}.

\bibitem{Nerling_2006}
F.~Nerling, J.~Blümer, R.~Engel and M.~Risse, \emph{Universality of electron distributions in high-energy air showers—description of cherenkov light production}, \href{https://doi.org/10.1016/j.astropartphys.2005.09.002}{\emph{Astroparticle Physics} {\bfseries 24} (2006) 421–437}.

\bibitem{Giller:2004cf}
M.~Giller, G.~Wieczorek, A.~Kacperczyk, H.~Stojek and W.~Tkaczyk, \emph{{Energy spectra of electrons in the extensive air showers of ultra-high energy}}, \href{https://doi.org/10.1088/0954-3899/30/2/009}{\emph{J. Phys. G} {\bfseries 30} (2004) 97}.

\bibitem{Conceicao:2015toa}
R.~Concei\c{c}\~ao, S.~Andringa, F.~Diogo and M.~Pimenta, \emph{{The average longitudinal air shower profile: exploring the shape information}}, \href{https://doi.org/10.1088/1742-6596/632/1/012087}{\emph{J. Phys. Conf. Ser.} {\bfseries 632} (2015) 012087}.

\bibitem{LIPARI2009309}
P.~Lipari, \emph{Universality of cosmic ray shower development}, \href{https://doi.org/https://doi.org/10.1016/j.nuclphysbps.2009.09.060}{\emph{Nuclear Physics B - Proceedings Supplements} {\bfseries 196} (2009) 309}.

\bibitem{muonUniv1}
L.~Cazon, R.~Conceição, M.~Pimenta and E.~Santos, \emph{A model for the transport of muons in extensive air showers}, \href{https://doi.org/https://doi.org/10.1016/j.astropartphys.2012.05.017}{\emph{Astroparticle Physics} {\bfseries 36} (2012) 211}.

\bibitem{muonUniv2}
M.~Ave, M.~Roth and A.~Schulz, \emph{A generalized description of the time dependent signals in extensive air shower detectors and its applications}, \href{https://doi.org/https://doi.org/10.1016/j.astropartphys.2017.01.003}{\emph{Astroparticle Physics} {\bfseries 88} (2017) 46}.

\bibitem{muonUniv3}
M.~Ave, R.~Engel, M.~Roth and A.~Schulz, \emph{A generalized description of the signal size in extensive air shower detectors and its applications}, \href{https://doi.org/https://doi.org/10.1016/j.astropartphys.2016.11.008}{\emph{Astroparticle Physics} {\bfseries 87} (2017) 23}.

\bibitem{Cazon_2023}
L.~Cazon, R.~Conceição and F.~Riehn, \emph{Universality of the muon component of extensive air showers}, \href{https://doi.org/10.1088/1475-7516/2023/03/022}{\emph{Journal of Cosmology and Astroparticle Physics} {\bfseries 2023} (2023) 022}.

\bibitem{muonUniv}
P.R.~Blake and W.F.~Nash, \emph{Muons in extensive air showers. i. the lateral distribution of muons}, \href{https://doi.org/10.1088/0954-3899/21/1/013}{\emph{Journal of Physics G: Nuclear and Particle Physics} {\bfseries 21} (1995) 129}.

\bibitem{PierreAuger:2018gfc}
{\scshape Pierre Auger} collaboration, \emph{{Measurement of the average shape of longitudinal profiles of cosmic-ray air showers at the Pierre Auger Observatory}}, \href{https://doi.org/10.1088/1475-7516/2019/03/018}{\emph{JCAP} {\bfseries 03} (2019) 018} [\href{https://arxiv.org/abs/1811.04660}{{\ttfamily 1811.04660}}].

\bibitem{MARTA}
P.~Abreu et~al., \emph{{MARTA: a high-energy cosmic-ray detector concept for high-accuracy muon measurement}}, \href{https://doi.org/10.1140/epjc/s10052-018-5820-2}{\emph{Eur. Phys. J. C} {\bfseries 78} (2018) 333} [\href{https://arxiv.org/abs/1712.07685}{{\ttfamily 1712.07685}}].

\bibitem{Slupecki:2888741}
M.~Slupecki, D.~Aguglia, Y.~Aguiar, R.~Alemany~Fernandez, T.~Argyropoulos, R.A.~Barlow et~al., \emph{{LHC Oxygen Run Preparation in the CERN Injector Complex}}, .

\bibitem{CORSIKA}
D.~Heck, J.~Knapp, J.~Capdevielle, G.~Schatz and T.~Thouw, \emph{A monte carlo code to simulate extensive air showers}, {\emph{Report FZKA} {\bfseries 6019} (1998) }.

\bibitem{Geant4_2006}
J.~Allison et~al.{\emph{IEEE Transactions on Nuclear Science} {\bfseries 53 No. 1} (2006) 270}.

\bibitem{Geant4_2016}
J.~Allison et~al.{\emph{Nuclear Instruments and Methods in Physics Research A} {\bfseries 835} (2016) 186}.

\bibitem{Offline}
S.~Argirò, S.~Barroso, J.~Gonzalez, L.~Nellen, T.~Paul, T.~Porter et~al., \emph{The offline software framework of the pierre auger observatory}, \href{https://doi.org/10.1016/j.nima.2007.07.010}{\emph{Nuclear Instruments and Methods in Physics Research Section A: Accelerators, Spectrometers, Detectors and Associated Equipment} {\bfseries 580} (2007) 1485–1496}.

\bibitem{PierreAuger:2015eyc}
{\scshape Pierre Auger} collaboration, \emph{{The Pierre Auger Cosmic Ray Observatory}}, \href{https://doi.org/10.1016/j.nima.2015.06.058}{\emph{Nucl. Instrum. Meth. A} {\bfseries 798} (2015) 172} [\href{https://arxiv.org/abs/1502.01323}{{\ttfamily 1502.01323}}].

\bibitem{PierreAuger:2021hun}
{\scshape Pierre Auger} collaboration, \emph{{The energy spectrum of cosmic rays beyond the turn-down around $10^{17}$ eV as measured with the surface detector of the Pierre Auger Observatory}}, \href{https://doi.org/10.1140/epjc/s10052-021-09700-w}{\emph{Eur. Phys. J. C} {\bfseries 81} (2021) 966} [\href{https://arxiv.org/abs/2109.13400}{{\ttfamily 2109.13400}}].

\bibitem{PierreAuger:2020flu}
{\scshape Pierre Auger} collaboration, \emph{{Studies on the response of a water-Cherenkov detector of the Pierre Auger Observatory to atmospheric muons using an RPC hodoscope}}, \href{https://doi.org/10.1088/1748-0221/15/09/P09002}{\emph{JINST} {\bfseries 15} (2020) P09002} [\href{https://arxiv.org/abs/2007.04139}{{\ttfamily 2007.04139}}].

\bibitem{PierreAuger:2005znw}
{\scshape Pierre Auger} collaboration, \emph{{Calibration of the surface array of the Pierre Auger Observatory}}, \href{https://doi.org/10.1016/j.nima.2006.07.066}{\emph{Nucl. Instrum. Meth. A} {\bfseries 568} (2006) 839} [\href{https://arxiv.org/abs/2102.01656}{{\ttfamily 2102.01656}}].

\bibitem{PierreAuger:2020gxz}
{\scshape Pierre Auger} collaboration, \emph{{Direct measurement of the muonic content of extensive air showers between $\mathbf { 2\times 10^{17}}$ and $\mathbf {2\times 10^{18}}~$eV at the Pierre Auger Observatory}}, \href{https://doi.org/10.1140/epjc/s10052-020-8055-y}{\emph{Eur. Phys. J. C} {\bfseries 80} (2020) 751}.

\end{thebibliography}\endgroup

\end{document}